\documentclass[journal,draftcls,onecolumn,12pt,twoside]{IEEEtran}
\usepackage{setspace}
\usepackage{textcomp}
\usepackage{stfloats}
\usepackage{url}
\usepackage{verbatim}
\usepackage[T1]{fontenc}
\usepackage{ifpdf}

\usepackage{cite}

\usepackage{graphicx}
\graphicspath{{../}}
\DeclareGraphicsExtensions{.pdf,.jpeg,.png,.eps}

\usepackage{enumitem}

\usepackage{amsthm}
\usepackage{algorithmic}
\usepackage{amsmath}
\usepackage{amsfonts}
\usepackage{amssymb}
\usepackage{mathrsfs}
\usepackage[cmintegrals]{newtxmath}
\usepackage{bm}

\newtheorem{theorem}{\textbf{Theorem}}
\newtheorem{proposition}{\textbf{Proposition}}
\newtheorem{definition}{\textbf{Definition}}
\newtheorem{example}{\textbf{Example}}

\usepackage[ruled,linesnumbered]{algorithm2e}
\usepackage{array}
\usepackage{multirow}
\usepackage[table,xcdraw]{xcolor}
\usepackage{booktabs,threeparttable}
\usepackage[caption=false,font=normalsize,labelfont=sf,textfont=sf]{subfig}

\begin{document}
\title{Polar Complexity: A New
\\Descriptive Complexity with Applications to \\Source and Joint Source-Channel Coding}
\author{Xinyuanmeng~Yao~\IEEEmembership{Member,~IEEE}
and Xiao~Ma~\IEEEmembership{Member,~IEEE}
\thanks{
Xinyuanmeng Yao is with the School of Cyber Science and Engineering and also with Ningbo Key Laboratory of Information Technology Application   Innovation and Security, Ningbo University of Technology, Ningbo 315211, China. She is currently a visiting scholar with the Department of Electrical and Electronic Engineering, Imperial College London, London SW7 2AZ, U.K. (e-mail: yaoxym@nbut.edu.cn).


Xiao Ma is with the School of Computer Science and Engineering, and also with Guangdong Province Key Laboratory of Information Security Technology, Sun Yat-sen University, Guangzhou 510006, China (e-mail: maxiao@mail.sysu.edu.cn). 

The corresponding author is Xiao Ma.}}
\maketitle
\begin{abstract}
This paper first presents a new approach to evaluating the descriptive complexity of finite-length binary sequences. Specifically, we investigate the sequence-wise recovery behavior induced by polar compression and successive cancellation decoding (SCD), and define the polar complexity of a sequence as the minimum polar-compression length~(PCL) required for its exact reconstruction. To compute the polar complexity efficiently, we further develop both a bisection-search algorithm and a low-complexity estimation method.

We then propose a polar-based two-stage source coding scheme, in which each source sequence is represented by its polar complexity followed by the corresponding polar-compressed sequence. The proposed scheme is strictly lossless and prefix-free. In addition, for BMSs, the normalized average compression length of the proposed scheme can asymptotically approach the source entropy under certain conditions. Simulation results further demonstrate that the scheme can operate without prior knowledge of the source statistics and remains robust across different source distributions.

Finally, we integrate the proposed polar source coding with polar channel coding to develop an adaptive double-polar joint source-channel coding~(JSCC) scheme, where the encoder and decoder share a predefined set of candidate PCLs to balance error performance and decoding complexity. We formulate the design of the candidate-PCL set as an optimization problem and solve it efficiently via dynamic programming. Simulation results show that the proposed adaptive double-polar JSCC scheme provides a flexible performance-complexity tradeoff and outperforms existing polar-code-based JSCC baselines.
\end{abstract}


\section{Introduction}
Efficiently describing messages is a fundamental problem in information theory. Shannon modeled a message as an output generated by a random source with known statistical characteristics and investigated the minimum average number of bits required to describe messages generated by that source. He further showed that the minimum achievable average description length is essentially characterized by the source entropy. Following Shannon's framework, a variety of source coding schemes have been developed, including Shannon coding~\cite{Shannon1948AMT}, Huffman coding~\cite{Huffman1952} and arithmetic coding~\cite{Rissanen1976}. This probabilistic formulation is well suited to communication systems in which sequences are generated according to a known or well-justified distribution. However, it is less directly applicable to individual sequences, for which an underlying probabilistic model may be unavailable or difficult to justify.

In contrast, Kolmogorov investigated the minimum number of bits required to generate an individual message. From an algorithmic perspective, he defined the Kolmogorov complexity of a message as the length of the shortest program capable of generating it~\cite{Kolmogorov1968ThreeAT}. Although universal by definition, Kolmogorov complexity is generally uncomputable and therefore does not directly yield practical description methods. This limitation has motivated the search for computable and efficient approaches to describing individual sequences. A natural line of research is to exploit the characteristics and regularities exhibited by sequences. One classical example is enumerative coding, which represents a finite-length binary sequence by its Hamming weight together with its lexicographic index within the corresponding type class~\cite{cover1999elements}. By associating sequence complexity with the progressive emergence of new patterns, Lempel and Ziv introduced the notion of Lempel-Ziv~(LZ) complexity in~\cite{lempel1976}, which subsequently gave rise to a family of LZ compression algorithms~\cite{Ziv1977universal,ziv1978compression}. Extending the exploitation of repeated structures in LZ algorithms, grammar-based compression methods were proposed in~\cite{Kieffer2000}, where a sequence is represented through a hierarchical generative grammar that recursively captures its repeated patterns. Another representative direction for describing individual sequences arises from the assumption that the sequence is generated by a linear system. In~\cite{Massey1969ShiftregisterSA}, Massey introduced the linear complexity of a sequence, defined as the length of the shortest linear feedback shift register~(LFSR) capable of generating it, thereby leading to an LFSR-based description method. Although computationally efficient, this method is inherently restricted to sequences whose structures can be effectively characterized by linear recurrences~\cite{Campbell1999LFSRsDN}. These examples illustrate a useful principle: once a tractable generative or recovery mechanism is specified, it naturally induces a computable description method, and consequently a corresponding notion of descriptive complexity, for individual sequences.

Polar coding offers a particularly appealing mechanism for this purpose. In~\cite{Arikan2009Polarization}, Ar{\i}kan introduced polar codes and proved that they can achieve the capacity of binary-input discrete memoryless channels~(BI-DMCs). Motivated by the duality between channel coding and source coding~\cite{Allard1972AlgebraicSource,Weiss1962CompressionCoding}, polar codes were later extended to fixed-length source coding. In such schemes, a polar-compression length~(PCL) is
prescribed in advance, and a source sequence is compressed into a sequence of this length based on polar compression and then reconstructed using successive cancellation decoding~(SCD). For BMSs, although not every source sequence can be recovered without error, the fixed-length polar source coding scheme achieves the source entropy with an asymptotically vanishing error probability~\cite{Arikan2010SourcePolarization}. Several extensions and refinements have since been developed, including those in~\cite{Korada2009PolarCodes,Cronie2010LosslessPolar,Cayci2013QaryPolar,Liu2025PolarCompression}. 
Moreover, fixed-length polar source coding can be concatenated with polar channel coding to form double-polar coding schemes for joint source-channel coding~(JSCC)~\cite{Dong2021DoublePolarBP,Dong2022DoublePolarJSCL}, and a number of enhanced variants have been investigated in~\cite{Zheng2021DoubleCRCPolar,Zheng2023DoublePAC,Wang2024PolarizationExtension}.

Although conventional fixed-length polar source coding is designed for stochastic sources, its underlying polar compression and SCD reconstruction procedure also provides a natural description mechanism for individual sequences. For a given sequence, it is therefore meaningful to ask for the minimum PCL that guarantees exact reconstruction under SCD. The resulting polar-compressed sequence then serves as a compact description of the original sequence.

In this paper, we adopt polar coding implemented over a binary balanced tree~(BBT) structure~\cite{Yao2022ABT}, referred to as BBT polar coding, as the underlying framework\footnote{Unless otherwise specified, we omit the qualifier ``BBT'' for brevity; all polar codes considered in this paper are based on the BBT construction.}. The BBT transform generalizes the original polar transform to arbitrary block lengths by recursively applying local polar transforms over a binary balanced tree. It preserves the polarization mechanism while removing the power-of-two length constraint, thereby enabling greater flexibility in practical implementations. The main contributions of this paper are summarized as follows.
\begin{itemize}
\item Based on the polar compression and SCD reconstruction procedure, we introduce a new descriptive complexity, termed polar complexity, defined as the minimum PCL required to guarantee exact reconstruction. To compute the polar complexity, we develop an efficient bisection-search algorithm and further propose a low-complexity estimation method.
\item We propose a polar-based two-stage source coding scheme in which each source sequence is represented by its polar complexity followed by the corresponding polar-compressed sequence. For a BMS, we prove that, when the source length is a power of two, the normalized average compression length of the proposed scheme asymptotically approaches the source entropy. Simulation results show that the scheme can be implemented without prior knowledge of the source statistics and remains robust
across a range of source distributions.
\item We further generalize the polar-based two-stage source coding and integrate it with the polar channel coding to develop an adaptive double-polar JSCC scheme. Our proposed JSCC scheme operates over a predefined set of candidate PCLs shared by the encoder and decoder. For each source sequence, the encoder selects an appropriate PCL, generates the corresponding polar-compressed sequence, and applies cyclic redundancy check~(CRC) encoding, zero-padding and adjustable polar channel encoding. The decoder performs hypothesis testing over the candidate PCLs and selects the final estimate according to the maximum a posteriori~(MAP) criterion. We also formulate the design of the candidate-PCL set as an optimization problem and solve it efficiently via dynamic programming, which enables a principled tradeoff between performance and complexity. Simulation results show that the proposed scheme can outperform existing polar-code-based JSCC baselines.
\end{itemize}

The remainder of this paper is organized as follows. Section~\ref{sec2} reviews BBT polar coding for channel coding. Section~\ref{sec3} introduces the polar-induced description, defines the polar complexity. It also presents an efficient method for exact computation, along with a low-complexity estimation approach. Section~\ref{sec4} presents the polar-based two-stage source coding scheme and proves its entropy-achieving property. Section~\ref{sec5} presents the adaptive double-polar JSCC scheme and the candidate-PCL set design method. Section~\ref{sec6} concludes the paper.

\section{Preliminaries}\label{sec2}
In this section, we briefly review the BBT polar coding scheme proposed in~\cite{Yao2022ABT} for channel coding, which serves as the underlying coding scheme in this work.
\subsection{BBT Structure}
\begin{figure*}[t]
\centering
\begin{minipage}[t]{0.38\textwidth}
\vskip-5cm
\centering
\includegraphics[width=0.7\textwidth]{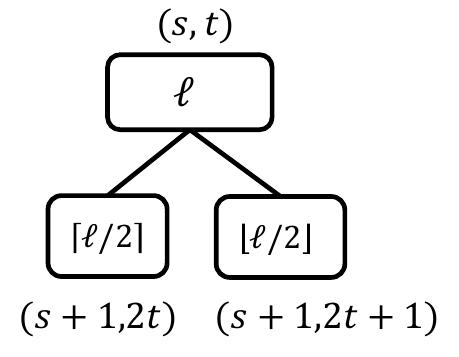}
\vskip1.5cm
\caption{The parent-child relationship in a BBT.}
\label{NodeRelationship}
\end{minipage}
\hfill
\begin{minipage}[t]{0.58\textwidth}
\centering
\includegraphics[width=\textwidth]{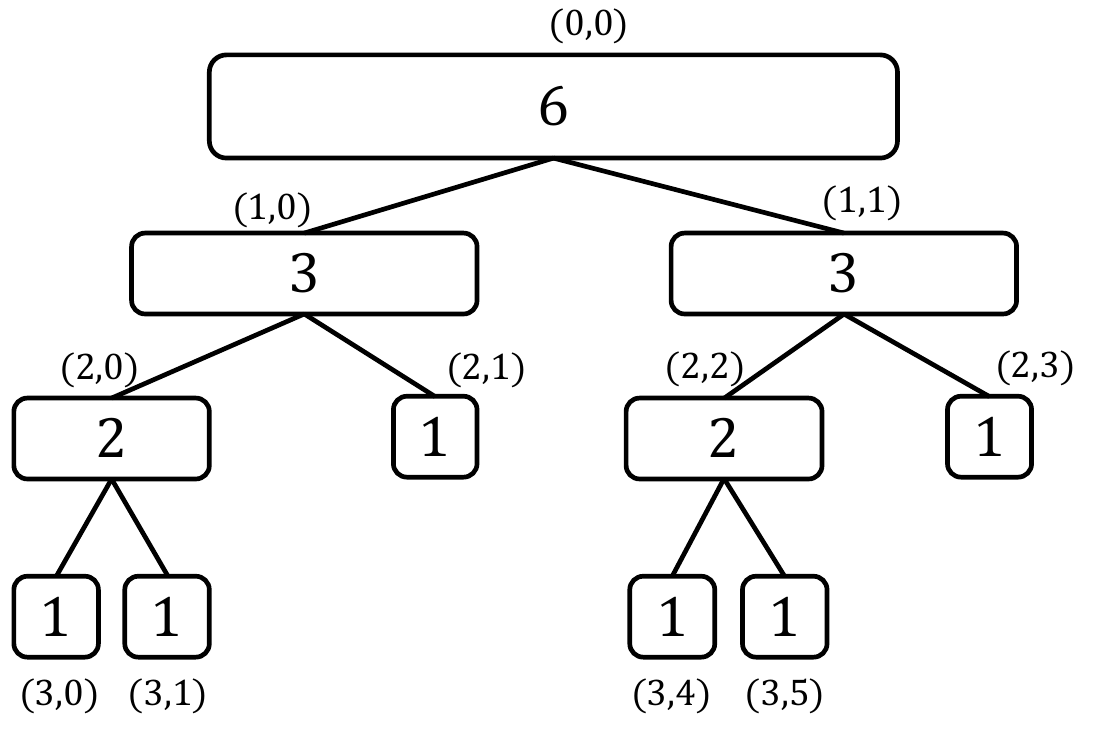}
\caption{The BBT structure for $N = 6$.}
\label{BBTStructureForN6}
\end{minipage}
\end{figure*}
A BBT is a binary tree in which, for every internal node, the heights of the left and right subtrees differ by at most one. The root node is located at level~$0$ and corresponds to a length~$N$. Each internal node of length $\ell \leq N$ (not necessarily a power of two) is split into two children: a left child of length $\lceil \ell/2 \rceil$ and a right child of length $\lfloor \ell/2 \rfloor$, where $\lceil \cdot \rceil$ and $\lfloor \cdot \rfloor$ denote the ceiling and floor operations, respectively. By construction, the heights of the left and right subtrees of every internal node differ by at most one. Each node is indexed by a pair $(s,t)$, where $s$ denotes the level and $t$ the horizontal position. The root node is indexed by $(0,0)$, and the left and right children of a node $(s,t)$ are indexed by $(s+1,2t)$ and $(s+1,2t+1)$, respectively. The parent-child relationship is illustrated in Fig.~\ref{NodeRelationship}.

Based on this recursive construction, a BBT can be generated for any length $N$ by iteratively splitting all nodes with $\ell \geq 2$. The resulting tree has $n+1$ levels and exactly $N$ leaf nodes, where $n=\lceil \log_2 N \rceil$. An example for $N=6$ is shown in Fig.~\ref{BBTStructureForN6}.
\subsection{BBT Polar Coding}
The objective of channel coding is to map a data sequence into a longer codeword to enable reliable reconstruction at the receiver. We now describe the implementation of the BBT polar coding scheme for this purpose.
\subsubsection{Encoding}
The encoding process is performed over a tree induced by the BBT structure, where each node of length $\ell$ is associated with a binary vector of length $\ell$. For a parent node $(s,t)$ of length $\ell$, let $\boldsymbol{v}^{(s,t)} \in \mathbb{F}_2^{\ell}$ denote its associated vector. This vector is computed from the vectors associated with its two child nodes, $(s+1,2t)$ and $(s+1,2t+1)$, whose lengths are $\lceil \ell/2 \rceil$ and $\lfloor \ell/2 \rfloor$, respectively, as
\begin{equation}
\boldsymbol{v}^{(s,t)}
=
\Big(
\boldsymbol{v}^{(s+1,2t)} \boxplus \boldsymbol{v}^{(s+1,2t+1)},\,
\boldsymbol{v}^{(s+1,2t+1)}
\Big),
\end{equation}
where ``$\boxplus$'' denotes a length-dependent bit-wise modulo-2 addition defined as follows.
\begin{itemize}
\item If $\lceil \ell/2 \rceil = \lfloor \ell/2 \rfloor$, then
\begin{equation}
\boldsymbol{v}^{(s+1,2t)} \boxplus \boldsymbol{v}^{(s+1,2t+1)}
=
\big(
v^{(s+1,2t)}_0 \oplus v^{(s+1,2t+1)}_0,\,
\ldots,\,
v^{(s+1,2t)}_{\lceil \ell/2 \rceil - 1} \oplus v^{(s+1,2t+1)}_{\lceil \ell/2 \rceil - 1}
\big).
\end{equation}
\item If $\lceil \ell/2 \rceil = \lfloor \ell/2 \rfloor + 1$, then
\begin{equation}
\boldsymbol{v}^{(s+1,2t)} \boxplus \boldsymbol{v}^{(s+1,2t+1)}
=
\big(
v^{(s+1,2t)}_0 \oplus v^{(s+1,2t+1)}_0,\,
\ldots,\,
v^{(s+1,2t)}_{\lceil \ell/2 \rceil - 2} \oplus v^{(s+1,2t+1)}_{\lceil \ell/2 \rceil - 2},\,
v^{(s+1,2t)}_{\lceil \ell/2 \rceil - 1}
\big).
\end{equation}
\end{itemize}

\begin{figure}[t]
\centering
\includegraphics[width=0.9\textwidth]{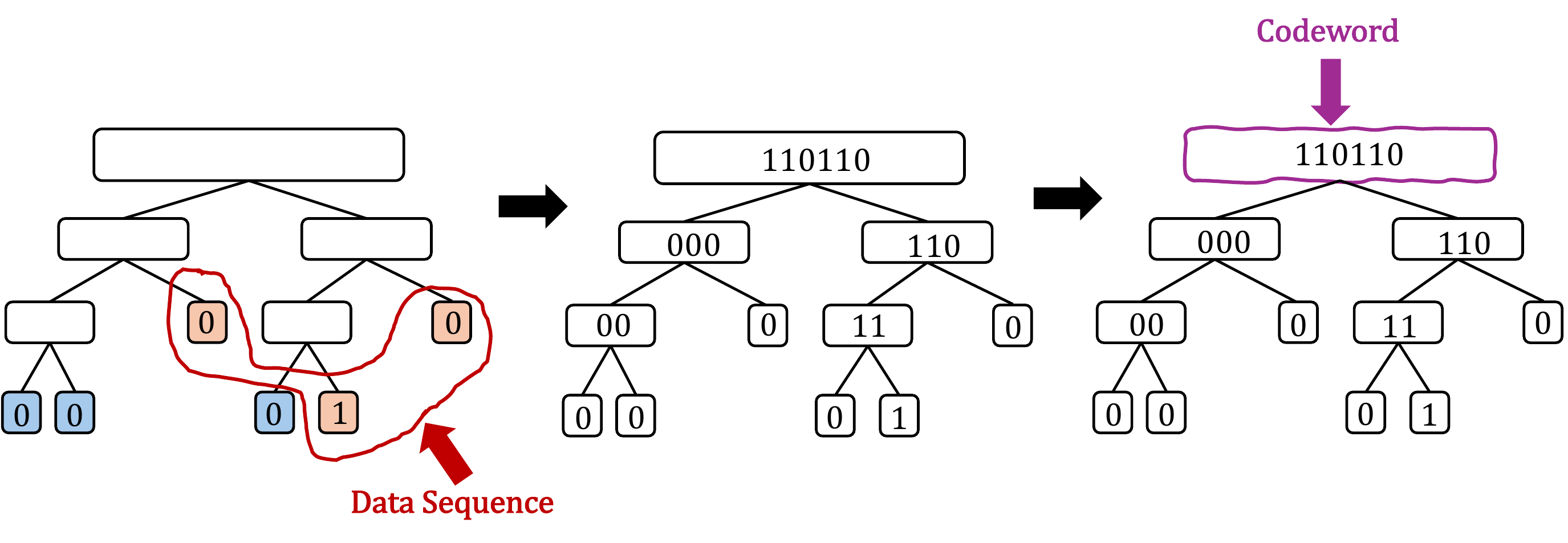}
\caption{Illustration of BBT polar encoding with $N=6$ and $K=3$, where blue leaf nodes denote frozen nodes and orange leaf nodes denote active nodes. The data sequence $(0,1,0)$ is encoded into the codeword $(1,1,0,1,1,0)$.}
\label{BBTPolarChannelEncoding}
\end{figure}

To encode a data sequence $\boldsymbol{u}\in\mathbb{F}_2^{K}$, the encoder assigns the $K$ data bits to the selected active leaf nodes, while setting the remaining $N-K$ frozen leaf nodes to zero. In this paper, an \emph{active} leaf node refers to a node whose value is unknown to the decoder, whereas a \emph{frozen} leaf node refers to a node whose value is known to the decoder. After initializing the bits associated with the leaf nodes, the vectors associated with all internal nodes are computed via a reverse level-order traversal. The resulting codeword is obtained at the root as $\boldsymbol{c}=\boldsymbol{v}^{(0,0)}\in\mathbb{F}_2^N$.An illustrative example with $N=6$ and $K=3$ is shown in Fig.~\ref{BBTPolarChannelEncoding}, where the data sequence $\boldsymbol{u}=(0,1,0)$ is encoded into the codeword $\boldsymbol{c}=(1,1,0,1,1,0)$.

\subsubsection{Decoding}
The SCD\footnote{To further improve decoding performance, SCD can be extended to successive cancellation list decoding~(SCLD) or CRC-aided SCLD, as shown in~\cite{Yao2022ABT}.} process is also performed over a tree induced by the BBT structure, where each node $(s,t)$ of length $\ell$ is associated with a log-likelihood ratio (LLR) vector $\boldsymbol{\alpha}^{(s,t)}$ and a hard-bit estimate (HBE) vector $\boldsymbol{\beta}^{(s,t)}$. Decoding starts from the root, where the LLRs are initialized based on the channel observations, and proceeds recursively according to the following rules.

\textbf{$\bullet$ LLR update for the left child:} Given the LLRs of a parent node, the LLRs of its left child are computed as follows. If $\ell$ is even,
\begin{align}
\alpha^{(s+1,2t)}_i
=
f\big(\alpha^{(s,t)}_i,\alpha^{(s,t)}_{\lceil\ell/2\rceil+i}\big),
\end{align}
for $0 \le i \le \lceil\ell/2\rceil-1$. If $\ell$ is odd,
\begin{align}
\alpha^{(s+1,2t)}_i=
\begin{cases}
f(\alpha^{(s,t)}_i,\alpha^{(s,t)}_{\lceil\ell/2\rceil+i}),
&0\le i\le\lceil\ell/2\rceil-2,\\
\alpha^{(s,t)}_i,
&i=\lceil\ell/2\rceil-1,
\end{cases}
\end{align}
where $f(a,b)=\ln\frac{1+e^{a+b}}{e^a+e^b}$.

\textbf{$\bullet$ LLR update for the right child:} After obtaining the HBE vector of the left child, the LLRs of the right child are updated as
\begin{align}
\alpha^{(s+1,2t+1)}_i
=
g\big(
\alpha^{(s,t)}_i,
\alpha^{(s,t)}_{\lceil\ell/2\rceil+i},
\beta^{(s+1,2t)}_i
\big),
\end{align}
for $0 \le i \le \lfloor\ell/2\rfloor-1$, where $g(a,b,c)=b+(-1)^c a$.

\textbf{$\bullet$ HBE update for the parent:} Once the HBE vectors of both children are available, the parent HBE vector is computed as
\begin{align}
\boldsymbol{\beta}^{(s,t)}
=
\big(
\boldsymbol{\beta}^{(s+1,2t)} \boxplus \boldsymbol{\beta}^{(s+1,2t+1)},
\boldsymbol{\beta}^{(s+1,2t+1)}
\big).
\end{align}

At the leaf nodes, the HBEs are determined as follows. For frozen nodes, the values are fixed to zero. For active nodes, decisions are made via threshold detection, i.e., $\beta=1$ if $\alpha \le 0$ and $\beta=0$ otherwise. After completing the traversal, the estimated data bits are extracted from the active leaf nodes.

\section{Polar Complexity}\label{sec3}
This section first presents a general fixed-length polar compression and SCD reconstruction procedure. This procedure induces a polar-based description of individual sequences, thereby motivating the notion of polar complexity. We further address the computation of polar complexity by developing an efficient exact algorithm as well as a low-complexity prediction method.
\subsection{Polar Compression and SCD Reconstruction}
Based on the BBT polar coding scheme introduced in Section~II-B, we consider the following fixed-length polar compression and SCD reconstruction procedure.

Let $\boldsymbol{s}=(s_0,s_1,\ldots,s_{K-1})\in\mathbb{F}_2^K$ be a binary sequence to be compressed under a prescribed PCL $L$. A BBT with root length $K$ is first constructed, and $L$ leaf nodes are selected as frozen nodes. The positions of these frozen nodes are assumed to be known to both the compressor and the reconstructor. In addition, an SCD parameter $\tau\in[0,1]$ is predetermined. At the reconstruction stage, the root LLRs of the BBT are initialized as $\alpha_i^{(0,0)}=\log\frac{1-\tau}{\tau}$, for $0\leq i\leq K-1$.

\begin{figure}[t]
\centering
\includegraphics[width=0.9\textwidth]{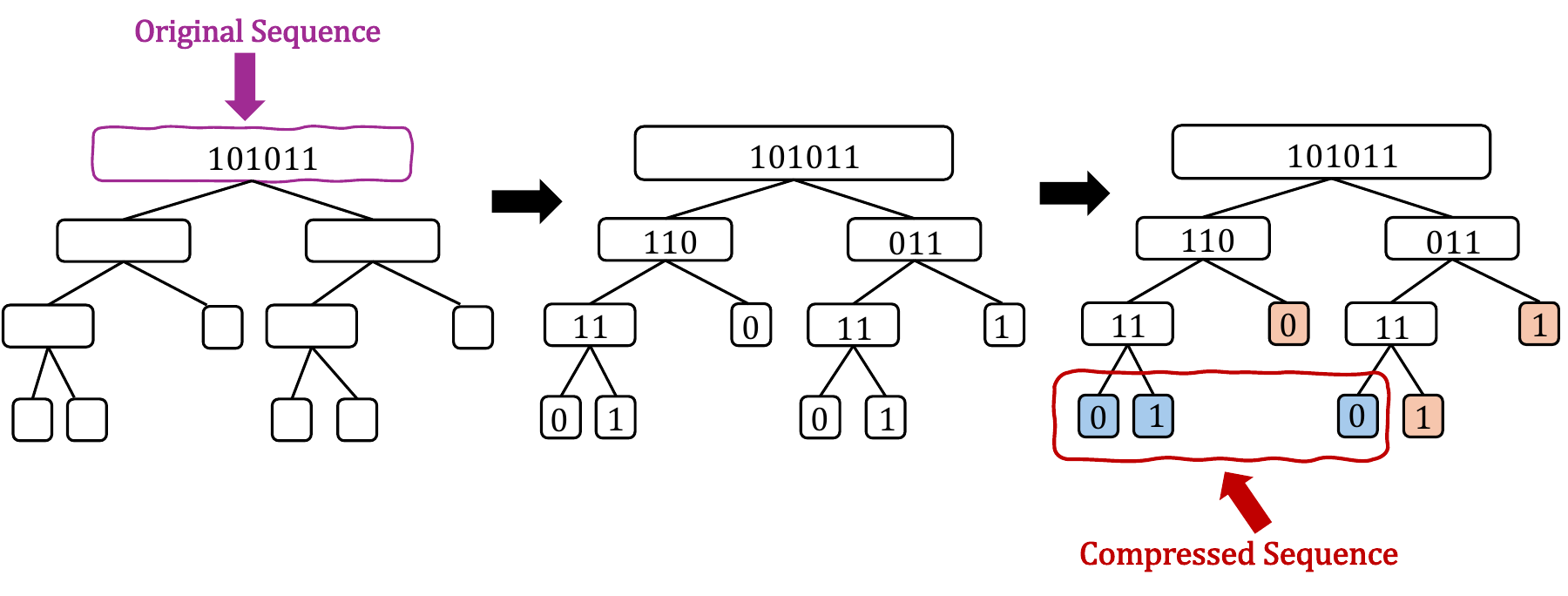}
\caption{Illustration of polar compression with $K=6$ and $L=3$, where blue leaf nodes denote frozen nodes and orange leaf nodes denote active nodes. The original sequence $(1,0,1,0,1,1)$ is compressed into the compressed sequence $(0,1,0)$.}
\label{BBTPolarSourceEncoding}
\end{figure}

During compression, the sequence $\boldsymbol{s}$ is assigned to the root node of the BBT. The vectors associated with all internal nodes are then computed by a level-order traversal according to the BBT polar transform. For the prescribed PCL $L$, the bits located at the $L$ selected frozen leaf nodes are extracted and ordered according to their left-to-right positions in the BBT. This yields an $L$-bit polar-compressed sequence, denoted by $\boldsymbol{d}=(d_0,d_1,\ldots,d_{L-1})\in\mathbb{F}_2^L$. An illustrative example with $K=6$ and $L=3$ is shown in
Fig.~\ref{BBTPolarSourceEncoding}, where
$\boldsymbol{s}=(1,0,1,0,1,1)$ is compressed into
$\boldsymbol{d}=(0,1,0)$. At the reconstruction stage, given $\boldsymbol{d}$, the values of the
corresponding $L$ selected frozen leaf nodes are set accordingly, and SCD is performed over the BBT to produce an estimate of the original source sequence, denoted by $\hat{\boldsymbol{s}}\in\mathbb{F}_2^K$.

In this procedure, only $L$ leaf-node values are revealed to the reconstructor. Consequently, only $2^L$ distinct sequences of length $K$ can be recovered. Therefore, unless $L=K$, recoverability cannot be guaranteed uniformly for all sequences in $\mathbb{F}_2^K$. This indicates that recoverability is inherently sequence-dependent: some sequences can be recovered from short compressed representations, whereas others require significantly larger $L$.

To formalize this behavior, we introduce the order in which frozen nodes are selected.
\begin{definition}[Frozen-Node Selection Sequence]
\label{def:nu}
Consider a BBT with root length $K$. A frozen-node selection sequence is a permutation of the leaf-node positions, denoted by
\begin{equation}
\boldsymbol{\nu}=(\nu_0,\nu_1,\ldots,\nu_{K-1}),
\end{equation}
where $\nu_i\in\{0,1,\ldots,K-1\}$ specifies the left-to-right position of the leaf node selected with zero-based priority index $i$. That is, for a PCL $L$, the leaf nodes at positions, $\nu_0,\nu_1,\ldots,\nu_{L-1}$, are selected as frozen nodes.
\end{definition}

\begin{definition}[Recoverability]
\label{def:recoverability}
For a sequence $\boldsymbol{s}\in\mathbb{F}_2^K$, let $\hat{\boldsymbol{s}}(L,\boldsymbol{\nu},\tau)\in\mathbb{F}_2^K$ denote the sequence reconstructed from the $L$-length polar-compressed sequence $\boldsymbol{d}(L,\boldsymbol{\nu},\tau)\in\mathbb{F}_2^L$ under the frozen-node selection sequence $\boldsymbol{\nu}$ and SCD parameter $\tau$. The sequence $\boldsymbol{s}$ is
said to be recoverable at PCL $L$ if $\hat{\boldsymbol{s}}(L,\boldsymbol{\nu},\tau)=\boldsymbol{s}$.
Equivalently, $\boldsymbol{d}(L,\boldsymbol{\nu},\tau)$ is called a recoverable polar-compressed sequence of $\boldsymbol{s}$ if it enables exact SCD reconstruction of $\boldsymbol{s}$.
\end{definition}

\begin{proposition}[Monotonicity of Recoverability]
\label{prop:mono}
If $\boldsymbol{s}$ is recoverable at PCL $L$, then it is also recoverable at any PCL $L'>L$.
\end{proposition}

\begin{proof}
Under the frozen-node selection sequence $\boldsymbol{\nu}$, the set of frozen
leaf nodes selected for PCL $L$ is a subset of that selected for any PCL
$L'>L$. Suppose that $\boldsymbol{s}$ is recoverable at PCL $L$. Then the SCD
reconstruction from $\boldsymbol{d}(L,\boldsymbol{\nu},\tau)$ produces the true source sequence $\boldsymbol{s}$, and hence all leaf-node decisions induced during this successful reconstruction are consistent with the true leaf-node values associated with $\boldsymbol{s}$.

Now consider the reconstruction procedure with PCL $L'>L$. Compared with the $L$-bit case, the reconstructor is additionally given the true values of the leaf nodes selected with priorities $L,L+1,\ldots,L'-1$. These additional revealed values coincide with the corresponding decisions that would have been made in the successful $L$-bit reconstruction. Therefore, replacing these already-correct decisions by their revealed true values does not change the successive decoding trajectory. The subsequent LLR updates and decisions remain consistent with those in the successful $L$-bit reconstruction. Consequently, we have $\hat{\boldsymbol{s}}(L',\boldsymbol{\nu},\tau)=\boldsymbol{s}$,
which proves the claim.
\end{proof}
\subsection{Polar Complexity}
Although the preceding polar compression and SCD reconstruction procedure allows arbitrary choices of the frozen-node selection sequence $\boldsymbol{\nu}$ and the SCD parameter $\tau$, we adopt a canonical configuration in this work.

Specifically, the frozen-node selection sequence $\boldsymbol{\nu}$ is constructed using the polarization-weight (PW) method~\cite{He2017,Yao2022ABT}, which provides a deterministic ordering aligned with the polarization structure. The SCD parameter is fixed to $\tau=0$\footnote{To remove the dependence on source priors, we fix the SCD parameter to $\tau = 0$, thereby making the resulting polar complexity a deterministic, sequence-wise measure independent of the source statistics. In fact, $\tau = 1$ could also be adopted. Owing to the binary symmetry of the polar transform, it yields behavior essentially symmetric to the $\tau = 0$ case.}, corresponding to $\alpha_i^{(0,0)} \to +\infty$. In practice, this is implemented by assigning sufficiently large LLR values, thereby enforcing deterministic initial decisions and eliminating prior uncertainty. Under this configuration, the compressed and reconstructed sequences are denoted simply by $\boldsymbol{d}(L)$ and $\hat{\boldsymbol{s}}(L)$, respectively.

By Proposition~\ref{prop:mono}, recoverability is monotonic in the PCL $L$. Therefore, for any sequence $\boldsymbol{s}$, there exists a PCL beyond which exact reconstruction is guaranteed. This observation motivates the following definition.
\begin{definition}[Polar Complexity]
\label{def:Lstar}
Consider the polar compression and SCD reconstruction procedure specified by the PW-based frozen-node selection sequence $\boldsymbol{\nu}$ and the SCD parameter $\tau=0$. For an individual sequence $\boldsymbol{s}\in\mathbb{F}_2^K$, its polar complexity is defined as the minimum PCL that guarantees exact reconstruction, namely,
\begin{equation}
L^\star(\boldsymbol{s})
\triangleq
\min\bigl\{
L\in\{1,\ldots,K\}:
\hat{\boldsymbol{s}}(L)=\boldsymbol{s}
\bigr\},
\label{eq:Lstar_def}
\end{equation}
The corresponding compressed sequence $\boldsymbol{d}(L^\star(\boldsymbol{s}))$ is referred to as the polar-induced description of $\boldsymbol{s}$.
\end{definition}

A smaller value of $L^\star(\boldsymbol{s})$ indicates that $\boldsymbol{s}$ admits a shorter polar-induced description under the specified polar compression and SCD reconstruction procedure.  

\subsection{Exact Computation of Polar Complexity}
For a length-$K$ sequence $\boldsymbol{s}$, a straightforward way to compute $L^\star(\boldsymbol{s})$ is to test recoverability sequentially for $L=1,2,\ldots,K$ and identify the first PCL at which exact reconstruction is achieved. Although this exhaustive search is conceptually simple, it may require up to $K$ SCD trials for each sequence, which becomes computationally expensive when $K$ is large.

The monotonicity property in Proposition~\ref{prop:mono} enables a more
efficient exact computation. Define the function 
\begin{align}
g_{\boldsymbol{s}}(L)\triangleq\mathbf{1}\{\hat{\boldsymbol{s}}(L)=\boldsymbol{s}\},
\end{align} 
where $\mathbf{1}\{\cdot\}$ denotes the indicator function, which equals one if the event inside the braces occurs and equals zero otherwise. By Proposition~\ref{prop:mono}, $g_{\boldsymbol{s}}(L)$ is non-decreasing in $L$. Therefore,
$L^\star(\boldsymbol{s})$ is the smallest value of $L$
for which $g_{\boldsymbol{s}}(L)=1$, and can be identified via
bisection search. The resulting procedure is summarized in
Algorithm~\ref{alg:bisection}.

\begin{algorithm}[t]
\caption{Exact computation of $L^\star(\boldsymbol{s})$ via bisection search}
\label{alg:bisection}
\begin{algorithmic}[1]
\STATE \textbf{Input:} original sequence $\boldsymbol{s}\in\mathbb{F}_2^K$
\STATE Initialize $L_{\min}=1$ and $L_{\max}=K$
\WHILE{$L_{\min}<L_{\max}$}
\STATE Set $L=\left\lfloor (L_{\min}+L_{\max})/2 \right\rfloor$
\STATE Generate the $L$-bit polar-compressed sequence $\boldsymbol{d}(L)$ from $\boldsymbol{s}$
\STATE Reconstruct $\hat{\boldsymbol{s}}(L)$ from $\boldsymbol{d}^{(L)}$ by SCD
\IF{$\hat{\boldsymbol{s}}(L)=\boldsymbol{s}$}
\STATE Set $L_{\max}=L$
\ELSE
\STATE Set $L_{\min}=L+1$
\ENDIF
\ENDWHILE
\STATE \textbf{Output:} $L^\star(\boldsymbol{s})=L_{\min}$
\end{algorithmic}
\end{algorithm}

\subsection{Prediction of Polar Complexity}
Although the exact computation via bisection search reduces the number of SCD trials from linear to logarithmic in $K$, it still requires repeated compression and reconstruction attempts. We next develop a low-complexity prediction method that estimates the polar complexity directly from the sequence. The key idea is to identify the leaf-node positions that are locally unreliable under SCD and should therefore be revealed in order to ensure exact reconstruction.

To this end, we use genie-aided SCD (GA-SCD). In GA-SCD, the LLR of each leaf node is computed under the assumption that all preceding leaf-node decisions are set to their true values. Thus, GA-SCD removes the effect of error propagation from previous incorrect decisions and isolates the local reliability of each leaf node.

\begin{definition}[Critical Set]
\label{def:hardset}
For a sequence $\boldsymbol{s}\in\mathbb{F}_2^K$, let
$\Lambda_i$ denote the LLR of the $i$-th leaf node computed during GA-SCD. The critical set is defined as
\begin{align}
\mathcal{C}(\boldsymbol{s})
\triangleq
&\Bigl\{i:\Lambda_i=0\Bigr\}
\nonumber\\
&\cup
\Bigl\{i:\Lambda_i\neq 0,
\operatorname{sgn}(\Lambda_i)
\text{is inconsistent with the true value of the $i$-th leaf node}
\Bigr\}.
\label{eq:hardset}
\end{align}
Here, $\operatorname{sgn}(\cdot)$ denotes the sign function.
\end{definition}

The critical set $\mathcal{C}(\boldsymbol{s})$ consists of leaf-node positions that are either ambiguous or would be incorrectly decided even under GA-SCD, where all preceding decisions are assumed to be correct. These positions are essential for exact reconstruction: if they are not revealed, they may induce local decision errors, whereas revealing them eliminates the corresponding sources of unreliability. All critical positions are covered once the selected frozen-node set includes $\mathcal{C}(\boldsymbol{s})$. Consequently, the required PCL is determined by the highest selection priority among the elements of $\mathcal{C}(\boldsymbol{s})$, which leads to the following prediction of polar complexity.

\begin{definition}[Predicted Polar Complexity]
\label{def:Lhat}
For a sequence $\boldsymbol{s}\in\mathbb{F}_2^K$ with critical set $\mathcal{C}(\boldsymbol{s})$, the predicted polar complexity is defined as
\begin{equation}
\widebar{L^{\star}}(\boldsymbol{s})
\triangleq
\begin{cases}
1+\displaystyle\max_{i\in\mathcal{C}(\boldsymbol{s})}
\boldsymbol{\nu}^{-1}(i),
& \mathcal{C}(\boldsymbol{s})\neq\varnothing,\\[1ex]
1,
& \mathcal{C}(\boldsymbol{s})=\varnothing,
\end{cases}
\label{eq:Lhat_def}
\end{equation}
where $\boldsymbol{\nu}^{-1}(i)$ denotes the zero-based selection priority of the $i$-th leaf node, i.e., $\nu_{\boldsymbol{\nu}^{-1}(i)}=i$.
\end{definition}

The computation of $\widebar{L^{\star}}(\boldsymbol{s})$ requires only one GA-SCD pass followed by a maximization over the critical set, thereby avoiding the repeated decoding trials required by exact bisection search. The predicted
value serves as a low-complexity surrogate for the exact polar complexity
$L^\star(\boldsymbol{s})$.

Moreover, since the predicted PCL ensures that all leaf nodes that are locally ambiguous or locally misdecoded under GA-SCD are revealed, it provides a conservative estimate of the shortest recoverable PCL. In particular, the prediction satisfies
\begin{equation}
\widebar{L^{\star}}(\boldsymbol{s})
\geq
L^\star(\boldsymbol{s}).
\label{eq:Lhat_geq_Lstar}
\end{equation}
The possible overestimation is mainly caused by zero-LLR positions, which are treated as critical in $\mathcal{C}(\boldsymbol{s})$ even though they may be resolved correctly by the actual SCD tie-breaking rule.

\section{Polar-Based Two-Stage Source Coding}\label{sec4}
In this section, we consider a general lossless source coding problem and propose a polar-based two-stage source coding scheme.
\subsection{Problem Statement}
Consider a BMS $S$ over $\mathbb{F}_2$ with distribution $P_S(1)=\theta$ and $P_S(0)=1-\theta$ for $0<\theta<1/2$. A general lossless source coding scheme can be described by a pair of mappings
\begin{equation}
f: \mathbb{F}_2^K \to \mathbb{F}_2^*, \qquad
g: \mathbb{F}_2^* \to \mathbb{F}_2^K,
\end{equation}
where $f$ and $g$ denote the encoder and decoder, respectively, and $\mathbb{F}_2^*$ represents the set of all finite-length binary sequences. Under this framework, a length-$K$ source sequence is compressed to a (non-empty) binary sequence of variable length. We require the pair $(f,g)$ satisfying the following conditions:
\begin{enumerate}
\item The encoder $f$ is injective and $g=f^{-1}$;
\item The range of $f$ is prefix-free, so that any concatenation of compressed
sequences can be uniquely and instantaneously decomposed.
\end{enumerate}
The first condition ensures strict losslessness, and the second guarantees practical decodability in a non-terminating transmission setting, where the encoder $f$ is applied successively to consecutive length-$K$ source sequences. The performance of the scheme is measured by the average compression length (ACL)
\begin{align}
\mathrm{ACL}\triangleq\mathbb{E}\bigl[|f(\boldsymbol{S}^K)|\bigr],
\end{align}
and the normalized ACL~(NACL)
\begin{align}
\mathrm{NACL}\triangleq \frac{1}{K}\,\mathbb{E}\bigl[|f(\boldsymbol{S}^K)|\bigr].
\end{align}
Here, $\mathbb{E}\bigl[\cdot\bigr]$ denotes the mathematical expectation, and $|f(\boldsymbol{S}^K)|$ represents the length of the compressed sequence $f(\boldsymbol{S}^K)$.
\subsection{Polar-Based Two-Stage Source Coding}
For each individual finite-length sequence, its polar complexity and the corresponding polar-induced description can be efficiently determined. Motivated by this observation, we propose a polar-based two-stage source coding scheme. Specifically, for a given source length $K$, the encoding and decoding procedures are described as follows.
\subsubsection{Encoding}
For a source sequence $\boldsymbol{s}\in\mathbb{F}_2^K$, the encoder first determines its polar complexity $L^{\star}(\boldsymbol{s})$ and accordingly obtains its polar-induced description
$\boldsymbol{d}_{\boldsymbol{s}}\in\mathbb{F}_2^{L^{\star}(\boldsymbol{s})}$.
The final source codeword is given by
$\bigl(\boldsymbol{b}_{\boldsymbol{s}},\boldsymbol{d}_{\boldsymbol{s}}\bigr)$,
where the header $\boldsymbol{b}_{\boldsymbol{s}}$ is the $k$-length binary representation of
$L^{\star}(\boldsymbol{s})-1$ with
$k\triangleq\lceil \log_2 K \rceil$.
\subsubsection{Decoding}
Upon receiving the source codeword
$\bigl(\boldsymbol{b}_{\boldsymbol{s}},\boldsymbol{d}_{\boldsymbol{s}}\bigr)$, the decoder first reads the $k$-bit header $\boldsymbol{b}_{\boldsymbol{s}}$ and recovers
$L^{\star}(\boldsymbol{s})$. It then reads the following $L^{\star}(\boldsymbol{s})$ bits, namely
$\boldsymbol{d}_{\boldsymbol{s}}$, and applies the SCD to reconstruct $\boldsymbol{s}$.

By construction, the proposed scheme is variable-length and strictly lossless. Moreover, since the fixed-length header uniquely specifies the length of the subsequent polar-induced description, the resulting code is prefix-free. Hence, the decoder can identify the endpoint of each source codeword immediately after reading the header.

Note that the proposed polar-based two-stage source coding scheme is fully determined by the configuration presented in Subsection~\ref{sec3}-B, in which the frozen-node selection sequence $\boldsymbol{\nu}$ is constructed using the PW method and the SCD parameter is fixed as $\tau=0$. This configuration is completely independent of the source distribution; hence, the proposed scheme can operate without any prior knowledge of the source statistics.

\subsection{Entropy-Achieving Property}
In this subsection, we analyze the asymptotic performance of the proposed
polar-based two-stage source coding scheme when the source distribution is known. To establish the entropy-achieving property, we consider a modified configuration in which both the frozen-node selection sequence and the SCD parameter are allowed
to depend on the source distribution.

Consider a BMS $S$ over $\mathbb{F}_2$ with $P_S(1)=\theta$ and $P_S(0)=1-\theta$,
where $0<\theta<1/2$. Let $\boldsymbol{S}^K=(S_0,S_1,\ldots,S_{K-1})$ be drawn i.i.d. from $S$, and let $\boldsymbol{U}^K=(U_0,U_1,\ldots,U_{K-1})$ denote its polar transform result. Specifically, $\boldsymbol{S}^K$ is assigned to the root of the BBT, and the vectors associated with all other nodes are computed recursively according to the BBT polar transform. The $i$-th component of $\boldsymbol{U}^K$ corresponds to the bit value at the $i$-th leaf node and is associated with a source Bhattacharyya parameter, defined as
\begin{align}
Z(U_i|U_0^{i-1},\boldsymbol{S}^K)
\triangleq
2\sum_{u_0^{i-1},\boldsymbol{s}^K}
P_{U_0^{i-1},\boldsymbol{S}^K}(u_0^{i-1},\boldsymbol{s}^K)
\sqrt{
P_{U_i|U_0^{i-1},\boldsymbol{S}^K}(0|u_0^{i-1},\boldsymbol{s}^K)
P_{U_i|U_0^{i-1},\boldsymbol{S}^K}(1|u_0^{i-1},\boldsymbol{s}^K)
},
\end{align}
where $U_0^{i-1}=(U_0,U_1,\ldots,U_{i-1})$. Based on the source Bhattacharyya parameters, we construct the frozen-node selection sequence $\boldsymbol{\nu}$ by ordering the indices in decreasing order of
$Z(U_i|U_0^{i-1},\boldsymbol{S}^K)$. The SCD parameter is set to $\tau=\theta$, which matches the true source distribution.

Under this modified configuration, for any sequence $\boldsymbol{s} \in \mathbb{F}_2^K$, let $L_{\theta}^{\star}(\boldsymbol{s})$ denote its minimum recoverable PCL. At PCL $L$, the corresponding polar-compressed and reconstructed sequences are denoted by $\boldsymbol{d}_{\theta}(L)$ and $\hat{\boldsymbol{s}}_{\theta}(L)$, respectively\footnote{The subscript $\theta$ emphasizes the dependence on the source distribution.}. In addition, we restrict the source length to powers of two. This setting is standard in the analysis of polar source coding and allows us to directly invoke classical results on conventional fixed-length polar source coding.

\begin{theorem}[\cite{Arikan2010SourcePolarization} ]
\label{thm:fixed_length_source}
Consider a BMS $S$ over $\mathbb{F}_2$ with $P_S(1)=\theta$ and $P_S(0)=1-\theta$, whose entropy is $H(\theta)$,
where $0<\theta<1/2$. For any compression rate $R>H(\theta)$, there exists a sequence of polar compression and SCD
reconstruction schemes with blocklength $K$ and PCL $L_K=\lceil K R\rceil$,
such that the reconstruction error probability satisfies
$\Pr\{\hat{\boldsymbol{S}}_{\theta}^K(L_K)\neq \boldsymbol{S}_{\theta}^K(L_K)\}\to 0$ as $K\to\infty$.
\end{theorem}

We are now ready to establish that the proposed polar-based two-stage source coding scheme is entropy-achieving.

\begin{theorem}
\label{thm:entropy_achieving}
Consider a BMS $S$ over $\mathbb{F}_2$ with $P_S(1)=\theta$ and $P_S(0)=1-\theta$, where $0<\theta<1/2$, and entropy $H(\theta)$. Under the modified configuration in which the frozen-node selection sequence $\boldsymbol{\nu}$ is determined by the source Bhattacharyya parameters and the SCD parameter is set to $\tau=\theta$, there exists a sequence of polar-based two-stage coding schemes with encoders $\{f_K\}$ such that
\begin{align}
\lim_{K\to\infty}
\frac{1}{K}\mathbb{E}\bigl[|f_K(\boldsymbol{S}^K)|\bigr]
=
H(\theta),
\end{align}
where $\boldsymbol{S}^K=(S_0,S_1,\ldots,S_{K-1})$ is drawn i.i.d. from the BMS $S$.
\end{theorem}

\begin{proof}
For any realization $\boldsymbol{s} \in \mathbb{F}_2^K$, the encoder outputs two parts, the binary representation of $L_\theta^{\star}(\boldsymbol{s})-1$ of length $\log_2 K$, and the corresponding polar-compressed sequence $\boldsymbol{d}_{\theta}(L_\theta^{\star}(\boldsymbol{s}))$ of length $L_\theta^{\star}(\boldsymbol{s})$. Hence, the total length of $f_K(\boldsymbol{s})$ is
\begin{equation}
|f_K(\boldsymbol{s})|
= \log_2 K + L_\theta^{\star}(\boldsymbol{s}).
\end{equation}
Taking expectation, we obtain
\begin{equation}
\label{TotalLength}
\mathbb{E}\bigl[|f_K(\boldsymbol{S}^K)|\bigr]
= \log_2 K + \mathbb{E}\bigl[L_\theta^{\star}(\boldsymbol{S}^K)\bigr].
\end{equation}

We next establish matching lower and upper bounds.

\textit{Lower bound:}
Since the proposed scheme is prefix-free, the Kraft inequality for prefix codes implies that
\begin{equation}
\mathbb{E}\bigl[|f_K(\boldsymbol{S}^K)|\bigr]
\ge H(\boldsymbol{S}^K)
= K H(\theta),
\end{equation}
where the equality follows from the i.i.d. assumption. Consequently,
\begin{equation}
\liminf_{K \to \infty}
\frac{1}{K}
\mathbb{E}\bigl[|f_K(\boldsymbol{S}^K)|\bigr]
\ge H(\theta).
\end{equation}

\textit{Upper bound:}
Fix any $R > H(\theta)$. By Theorem~\ref{thm:fixed_length_source}, there exists a sequence of polar compression and SCD reconstruction schemes with $L_K = \lceil K R \rceil$ such that the reconstruction error probability
\begin{align}
P_e^{(K)} \triangleq
\Pr\{\hat{\boldsymbol{S}}_{\theta}^K(L_K) \neq \boldsymbol{S}^K\}
\to 0
\quad \text{as } K \to \infty.
\end{align}
By definition, $L_\theta^{\star}(\boldsymbol{s})$ is the minimum PCL that guarantees exact reconstruction of $\boldsymbol{s}$. Therefore, for a realization $\boldsymbol{s}$ such that 
$\hat{\boldsymbol{s}}_{\theta}(L_K) = \boldsymbol{s}$, it holds that
\begin{align}
L_\theta^{\star}(\boldsymbol{s}) \le L_K.
\end{align}
Moreover, for any realization $\boldsymbol{s}$, we have 
\begin{align}
L_\theta^{\star}(\boldsymbol{s}) \le K,
\end{align}
since revealing all leaf-node bits guarantees exact reconstruction. Consequently,
\begin{equation}
L_\theta^{\star}(\boldsymbol{S}^K)
\le L_K \mathbf{1}\{\hat{\boldsymbol{S}}_{\theta}^K(L_K) = \boldsymbol{S}^K\}
+ K \mathbf{1}\{\hat{\boldsymbol{S}}_{\theta}^K(L_K) \neq \boldsymbol{S}^K\}.
\end{equation}

Taking expectations yields
\begin{equation}
\mathbb{E}\bigl[L_\theta^{\star}(\boldsymbol{S}^K)\bigr]
\le L_K (1 - P_e^{(K)}) + K P_e^{(K)}
\le \lceil K R \rceil + K P_e^{(K)}.
\end{equation}
Substituting into \eqref{TotalLength} gives
\begin{equation}
\mathbb{E}\bigl[|f_K(\boldsymbol{S}^K)|\bigr]
\le \log_2 K + \lceil K R \rceil + K P_e^{(K)}.
\end{equation}
Dividing by $K$, we obtain
\begin{equation}
\frac{1}{K}
\mathbb{E}\bigl[|f_K(\boldsymbol{S}^K)|\bigr]
\le \frac{\log_2 K}{K} + R + P_e^{(K)}.
\end{equation}

Letting $K \to \infty$, and noting that $\frac{\log_2 K}{K} \to 0$ and $P_e^{(K)} \to 0$, we have
\begin{equation}
\limsup_{K \to \infty}
\frac{1}{K}
\mathbb{E}\bigl[|f_K(\boldsymbol{S}^K)|\bigr]
\le R.
\end{equation}
Since this holds for all $R > H(\theta)$, it follows that
\begin{equation}
\limsup_{K \to \infty}
\frac{1}{K}
\mathbb{E}\bigl[|f_K(\boldsymbol{S}^K)|\bigr]
\le H(\theta).
\end{equation}
Combining the lower and upper bounds completes the proof.
\end{proof}

\subsection{Simulation Results}
In this subsection, we first investigate the behavior of the polar complexity and then evaluate the proposed two-stage source coding scheme.
\subsubsection{Polar Complexity}
\begin{figure}[t]
\centering
\includegraphics[width=0.68\linewidth]{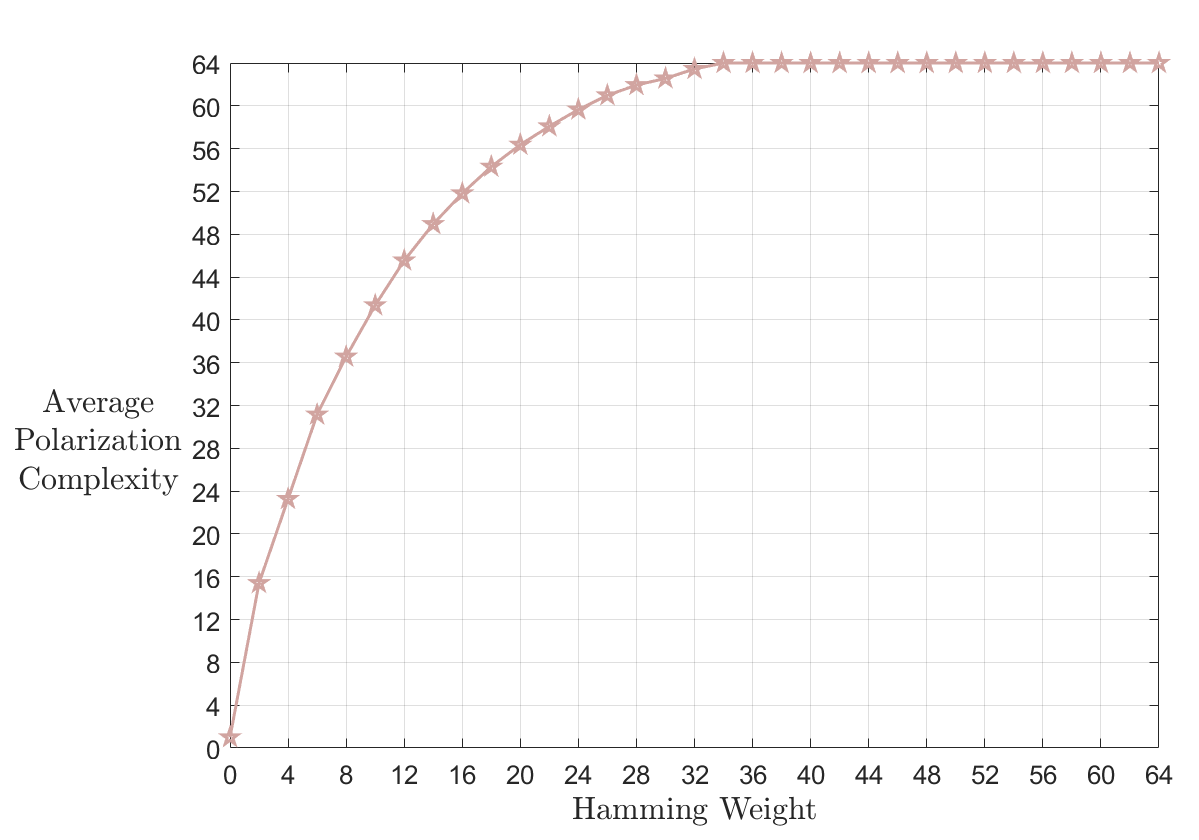}
\caption{Average polar complexity versus Hamming weight, where $K=64$.}
\label{AveragePolarizationComplexity_N64}
\end{figure}

We consider a sequence length of $K=64$ and evaluate the average polar complexity of binary sequences with varying Hamming weights. The results are presented in Fig.~\ref{AveragePolarizationComplexity_N64}.
As observed, the average polar complexity increases with the Hamming weight and eventually saturates. This behavior indicates that, under the considered polar compression and SCD reconstruction procedure specified by the PW-based frozen-node selection sequence and $\tau = 0$, sequences with smaller Hamming weights require significantly fewer frozen leaf-node bits to ensure exact reconstruction.

\subsubsection{Polar-Based Two-Stage Source Coding}
\paragraph{Illustration of the Coding Procedure}
We first illustrate our proposed source coding procedure through a simple example. Consider the set of all binary sequences of length $K=8$ and Hamming weight one. Table~\ref{tab:weight-one-example} shows the source codewords generated by the proposed polar-based two-stage source coding scheme, where, for each sequence $\boldsymbol{s}$, $\boldsymbol{b}_{\boldsymbol{s}}$ is the binary representation of its polar complexity and $\boldsymbol{d}_{\boldsymbol{s}}$ is its polar-induced description. For comparison, the table also presents the source codewords generated by enumerative coding. Although all sequences belong to the same Hamming-weight class, they exhibit different polar complexities. This illustrates that the proposed scheme captures structural differences induced by the polar transform, rather than relying solely on Hamming weight. 

\begin{table}[t]
\centering
\caption{Comparison between polar-based two-stage source coding \\and enumerative coding for binary sequences of length $8$ and Hamming weight one.}
\label{tab:weight-one-example}
\begin{tabular}{@{}ccllll@{}}
\toprule
$\boldsymbol{s}$ & Polar Complexity & \multicolumn{1}{c}{$\boldsymbol{b}_{\boldsymbol{s}}$} & \multicolumn{1}{c}{$\boldsymbol{d}_{\boldsymbol{s}}$} & \multicolumn{1}{c}{\begin{tabular}[c]{@{}c@{}}Polar-Based Two-Stage \\ Polar Source Coding\end{tabular}} & \multicolumn{1}{c}{Enumerative Coding} \\ \midrule
$10000000$       & 1                & 000                                                   & 1                                                     & 000,1                                                                                                    & 0001,111                               \\
$01000000$       & 2                & 001                                                   & 11                                                    & 001,11                                                                                                   & 0001,110                               \\
$00100000$       & 3                & 010                                                   & 101                                                   & 010,101                                                                                                  & 0001,101                               \\
$00010000$       & 3                & 011                                                   & 111                                                   & 011,1111                                                                                                 & 0001,100                               \\
$00001000$       & 4                & 100                                                   & 1001                                                  & 100,1001                                                                                                 & 0001,011                               \\
$00000100$       & 4                & 101                                                   & 1101                                                  & 101,1101                                                                                                 & 0001,010                               \\
$00000010$       & 4                & 110                                                   & 1011                                                  & 110,1011                                                                                                 & 0001,001                               \\
$00000001$       & 4                & 111                                                   & 1111                                                  & 111,1111                                                                                                 & 0001,000                               \\ \bottomrule
\end{tabular}
\end{table}

\paragraph{Performance over Different Sources}
\begin{figure}[t]
\centering
\begin{minipage}{0.48\linewidth}
    \centering
    \includegraphics[width=\linewidth]{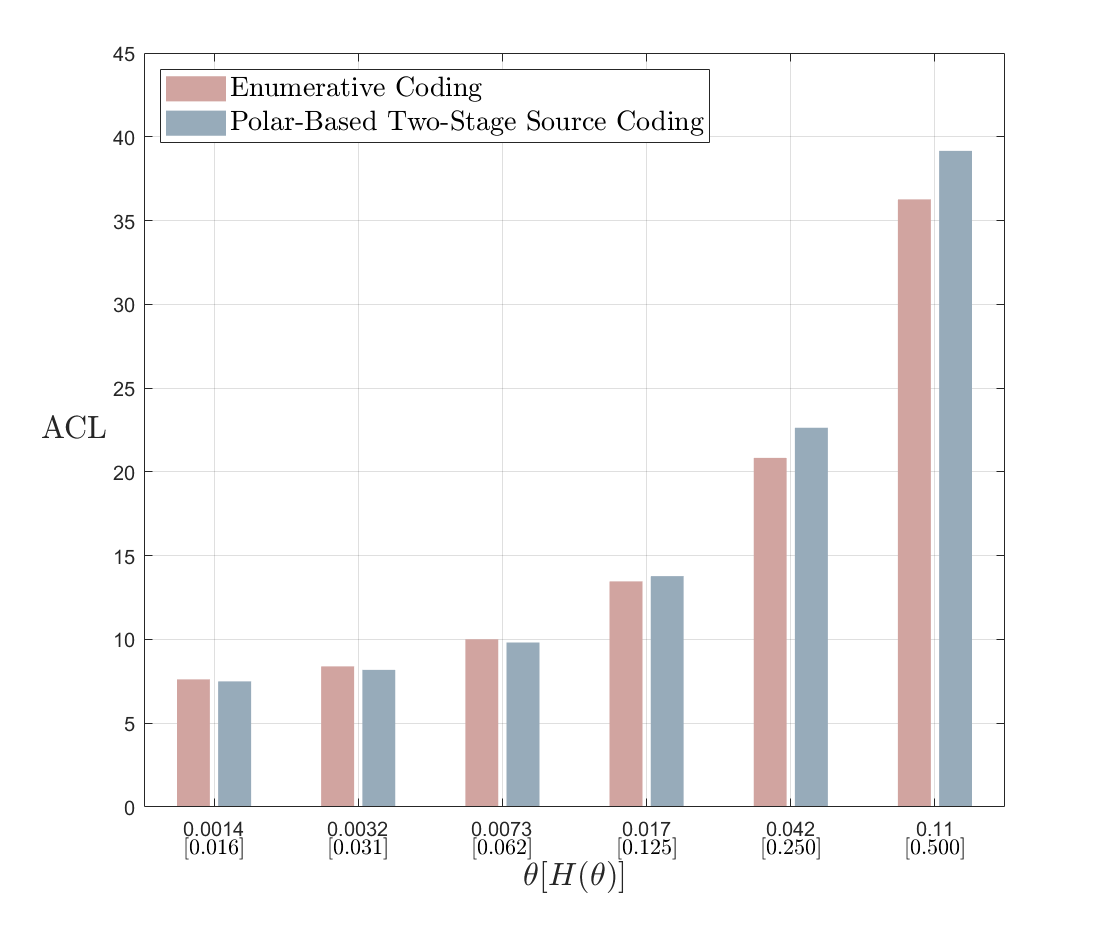}
    \caption{ACL performance of the polar-based two-stage source coding and enumerative coding versus the source entropy $H(\theta)$, where $K=64$.}
    \label{ACL_N64}
\end{minipage}
\hfill
\begin{minipage}{0.48\linewidth}
    \centering
    \includegraphics[width=\linewidth]{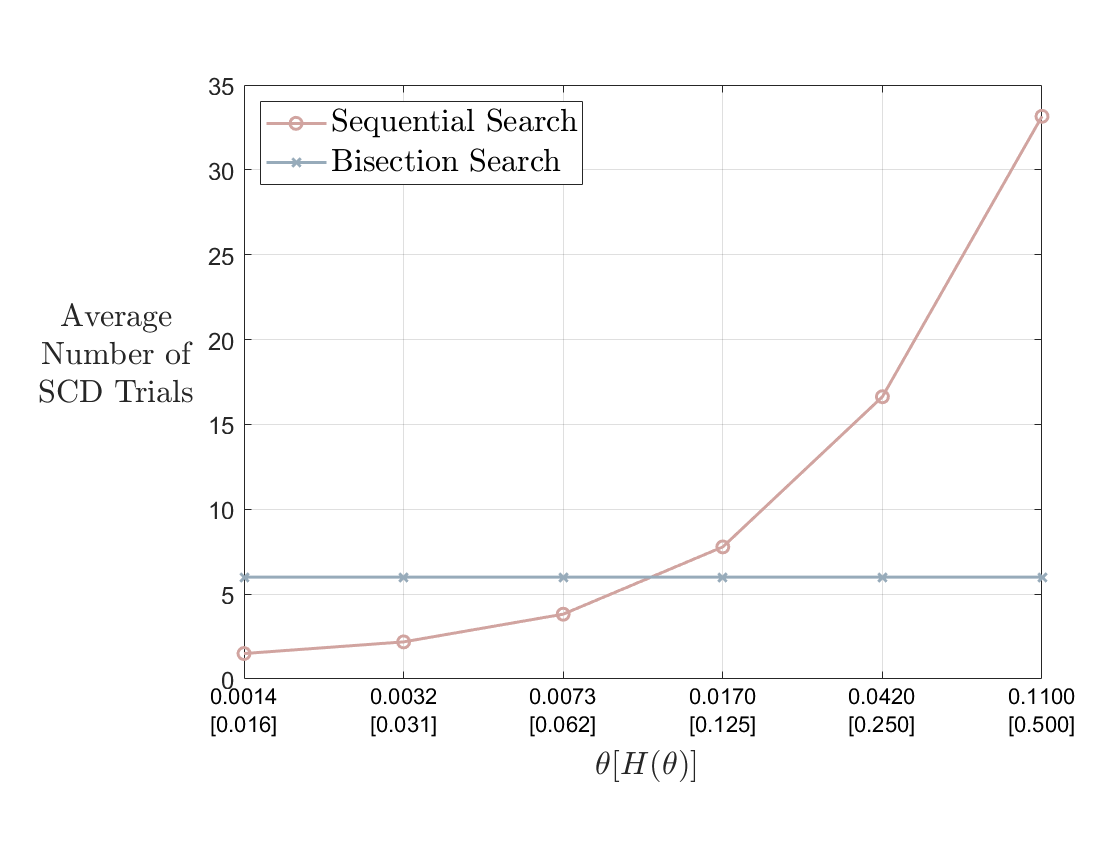}
    \caption{Average number of SCD trials required for computing the polar complexity versus the source entropy under the sequential search method and the bisection search method.}   \label{ThetaAverageNumberOfSCDTrials_N64}
\end{minipage}
\vskip-2em
\end{figure}
We next evaluate the compression performance of the proposed scheme over a range of BMSs with $\theta\in\{0.0014, 0.0032, 0.0073, 0.017, 0.042, 0.11\}$, corresponding approximately to $H(\theta)\in\{2^{-6},2^{-5},2^{-4},2^{-3},2^{-2},2^{-1}\}$. The ACL performance for blocklength $K=64$ is shown in Fig.~\ref{ACL_N64}.

As expected, the ACL increases monotonically with the source entropy, reflecting that higher-entropy sources require longer lossless descriptions on average. More importantly, in the short-blocklength regime, the proposed source coding scheme achieves performance comparable to that of enumerative coding over the considered entropy range. Enumerative coding typically incurs polynomial complexity, on the order of $\mathcal{O}(K^2)$, due to ranking and unranking operations over type classes. In contrast, the proposed polar-based two-stage scheme has significantly lower complexity, on the order of $\mathcal{O}(K\log^2 K)$, arising from the computation of polar complexity.

We also evaluate the computational cost of determining the polar complexity, measured by the average number of SCD trials. As shown in Fig.~\ref{ThetaAverageNumberOfSCDTrials_N64}, the bisection search method requires only on the order of $\log_2 K$ SCD trials and remains nearly insensitive to the source entropy. By contrast, the sequential search method requires an increasing number of SCD trials as the source entropy grows. These results confirm that the bisection search method substantially reduces the complexity of polar-complexity computation across a wide range of source statistics.
\paragraph{Comparison with Existing Schemes}
Finally, we compare the proposed scheme with existing polar-based source coding schemes~\cite{Cronie2010LosslessPolar,Liu2025PolarCompression}. Table~\ref{PerformanceComparison} presents the NACL results for \(K\in\{2^7,2^8,2^9,2^{10}\}\) and a BMS with \(\theta=0.11\). We also evaluate SCLDs with different list sizes as alternatives to SCD to improve compression performance.

As expected, the NACL decreases as the list size increases. In particular, with list size $8$, the proposed scheme achieves NACL performance comparable to that of existing polar-based source coding schemes. Moreover, we also consider the setting $\tau = \theta$. The resulting NACL is close to that obtained with $\tau = 0$, indicating that the proposed scheme is relatively insensitive to the choice of $\tau$ and exhibits a degree of universality.
\begin{table}[t]
\centering
\caption{NACL comparison of the polar-based two-stage source coding scheme \\and existing polar-based source coding schemes for $N\in\{2^7,2^8,2^9,2^{10}\}$ and a BMS with $\theta=0.11$.}
\label{PerformanceComparison}
\begin{tabular}{ccccc}
\toprule
$N$ & $2^7$ & $2^8$ & $2^9$ & $2^{10}$ \\
\midrule
\cite{Cronie2010LosslessPolar} & 0.606 & 0.586 & 0.580 & 0.568 \\
\cite{Liu2025PolarCompression} & 0.576 & 0.561 & 0.555 & 0.548 \\
Proposed Scheme with SCD, $\tau=0$ & 0.608 & 0.607 & 0.605 & 0.602\\ 
Proposed Scheme with SCLD(2), $\tau=0$ &0.590 & 0.586 & 0.585 & 0.584\\ 
Proposed Scheme with SCLD(4), $\tau=0$ &0.586 & 0.579 & 0.577 & 0.575\\ 
Proposed Scheme with SCLD(8), $\tau=0$ &0.585 & 0.579 & 0.573 & 0.571\\ 
Proposed Scheme with SCD, $\tau=\theta$ &
0.608 & 0.607 & 0.605 & 0.602\\
\bottomrule
\end{tabular}
\end{table}

\section{Adaptive Double-Polar JSCC}\label{sec5}
\subsection{Problem Statement}
Consider a BMS $S$ over $\mathbb{F}_2$ with distribution $P_S(1)=\theta$ and $P_S(0)=1-\theta$, where $0<\theta<1/2$, and a BI-DMC $W$ with transition probability $W_{Y|X}(y|x)$, $x \in \mathbb{F}_2$, $y \in \mathbb{R}$. A JSCC scheme is defined by a pair of mappings
\begin{align}
\phi:\mathbb{F}_2^K \rightarrow \mathbb{F}_2^N, 
\quad 
\psi: \mathbb{R}^N \rightarrow \mathbb{F}_2^K,
\end{align}
where $\phi$ and $\psi$ denote the encoder and decoder, respectively. A source sequence $\boldsymbol{s} \in \mathbb{F}_2^K$ is encoded into a codeword $\boldsymbol{x}=(x_0,x_1,\ldots,x_{N-1})\in\mathbb{F}_2^N$, which is transmitted over the channel $W$, resulting in a received sequence $\boldsymbol{y}=(y_0,y_1,\ldots,y_{N-1})\in\mathbb{R}^N$. The decoder operates on the LLRs
\begin{align}\label{LLR}
\lambda_i = \ln \frac{W_{Y|X}(y_i|0)}{W_{Y|X}(y_i|1)}, \quad i=0,1,\ldots,N-1,
\end{align}
together with the source parameter $\theta$, to produce an estimate $\hat{\boldsymbol{s}}\in\mathbb{F}_2^K$ of the original source sequence. The performance of the JSCC scheme is evaluated in terms of the frame error rate (FER), defined as $\mathrm{FER}\triangleq\Pr\{\boldsymbol{S}\neq\hat{\boldsymbol{S}}\}$.
\subsection{Proposed JSCC Scheme}
\begin{figure}
\centering
\includegraphics[width=0.8\linewidth]{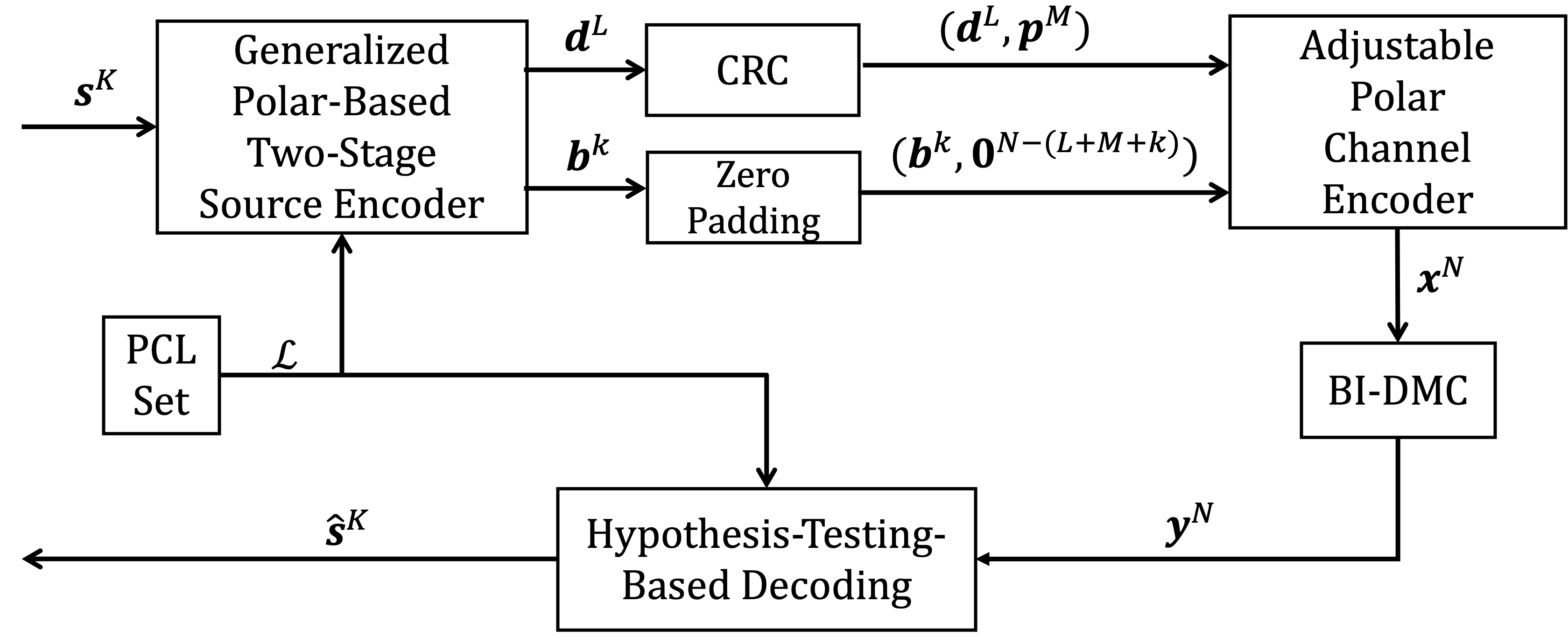}
\caption{Diagram of the proposed adaptive double-polar JSCC scheme.}
\label{DiagramOfAdaptiveDoubleBBTPolar}
\end{figure}
We propose an adaptive double-polar JSCC scheme that dynamically matches the compressibility of each source realization with the error-protection capability of the channel code. It builds upon the polar-based two-stage source coding scheme in Section~IV-B and the polar channel coding scheme in Section~II-B, with suitable modifications to form a unified JSCC framework, as illustrated in Fig.~\ref{DiagramOfAdaptiveDoubleBBTPolar}.

Within this framework, we define a finite PCL set
\begin{equation}
\mathcal{L}=\{l_0,l_1,\ldots,l_{m-1}\}\subseteq\{1,\ldots,K\}, 
\qquad
l_0<l_1<\cdots<l_{m-1}=K,
\end{equation}
which restricts the admissible PCL values to $m$ candidates known to both the encoder and the decoder. By restricting the possible PCLs to a finite candidate set, the proposed scheme reduces the encoding and decoding complexity while retaining adaptability to different source realizations.

\subsubsection{Encoding}
Given a source sequence $\boldsymbol{s}\in\mathbb{F}_2^K$ with predicted polar complexity $\overline{L^{\star}}(\boldsymbol{s})$, the encoder first selects its PCL according to
\begin{equation}\label{SelectedPCL_final}
L=\min\left\{l\in\mathcal{L}:l\geq \overline{L^{\star}}(\boldsymbol{s})\right\}.
\end{equation}
In other words, the encoder selects the smallest admissible PCL in $\mathcal{L}$ that is no smaller than the predicted polar complexity. Since the predicted polar complexity serves as an upper bound on the true polar complexity, this selection rule guarantees exact reconstruction in the absence of channel errors.

Once $L$ is determined, the encoding procedure consists of the following steps:
\begin{enumerate}[label=Step~\arabic*:,leftmargin=*]
\item Given the selected PCL $L$, a polar-compressed sequence $\boldsymbol{d}\in\mathbb{F}_2^L$ is generated from $\boldsymbol{s}$. An $M$-bit CRC sequence $\boldsymbol{p}\in\mathbb{F}_2^M$ is then computed from $\boldsymbol{d}$ and appended to it, forming
$(\boldsymbol{d},\boldsymbol{p})\in\mathbb{F}_2^{L+M}$.
\item The selected PCL $L$ is represented by a $k$-bit binary header $\boldsymbol{b}\in\mathbb{F}_2^k$, where $k=\lceil\log_2 K\rceil$, using the binary representation of $L-1$. This header is then padded with zeros to form
$(\boldsymbol{b},\boldsymbol{0})\in\mathbb{F}_2^{N-(L+M)}$.
\item An adjustable polar channel encoder is applied to generate the channel codeword
$\boldsymbol{x}\in\mathbb{F}_2^N$. The sequence $(\boldsymbol{d},\boldsymbol{p})$ is assigned to the $L+M$ active leaf nodes, while $(\boldsymbol{b},\boldsymbol{0})$ is assigned to the remaining
$N-(L+M)$ frozen leaf nodes.
\end{enumerate}
\begin{example}
\begin{figure}
\centering
\includegraphics[width=\linewidth]{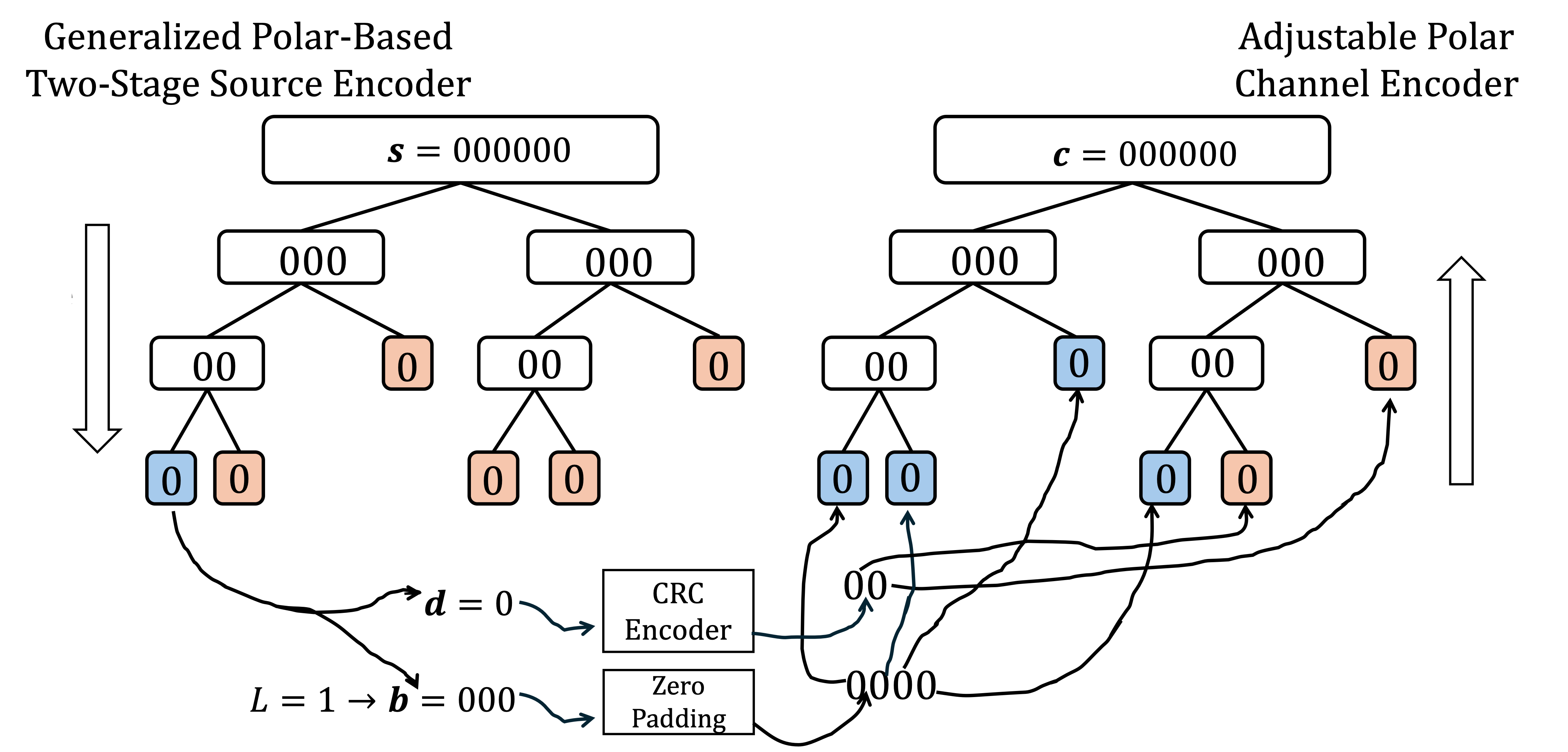}
\caption{Illustration of the encoding procedure for the adaptive double-polar JSCC scheme, where $(N,K,M,k)=(6,6,1,3)$.}
\label{AdaptiveDoubleBBTPolarEncoding}
\end{figure}

Fig.~\ref{AdaptiveDoubleBBTPolarEncoding} shows a simple example with $K=6$ and $N=6$. For the all-zero source sequence $\boldsymbol{s}=(0,0,0,0,0,0)$, suppose the selected PCL is $L=1$.

The generalized polar-based two-stage source encoder produces $\boldsymbol{d}=(0)$ along with a $3$-bit header $\boldsymbol{b}=(0,0,0)$, where $k=\lceil\log_2 6\rceil=3$. A one-bit CRC is appended to $\boldsymbol{d}$, yielding $(\boldsymbol{d},\boldsymbol{p})=(0,0)$. The header $\boldsymbol{b}$ is extended via zero padding to $(\boldsymbol{b},\boldsymbol{0})=(0,0,0,0)$. The polar channel encoder then selects two leaf nodes as active nodes, assigning $(\boldsymbol{d},\boldsymbol{p})$ to them, while mapping $(\boldsymbol{b},\boldsymbol{0})$ to the frozen nodes. After applying the BBT transform, the resulting codeword is $\boldsymbol{c}=(0,0,0,0,0,0)$.
\end{example}

\subsubsection{Decoding}
Upon receiving the channel output $\boldsymbol{y}\in\mathbb{R}^N$, the receiver first computes the corresponding LLR vector
$\boldsymbol{\lambda}\in\mathbb{R}^N$ according to~\eqref{LLR}. Since the selected PCL $L$ is not known a priori at the receiver, the decoder performs hypothesis testing over all candidates in $\mathcal{L}$.

For each candidate $l\in\mathcal{L}$, the decoder performs the following steps:
\begin{enumerate}[label=Step~\arabic*:,leftmargin=*]
\item The candidate PCL $l$ is represented by a $k$-bit binary header
$\tilde{\boldsymbol{b}}\in\mathbb{F}_2^k$ using the binary representation of $l-1$. This header is padded with zeros to form
$(\tilde{\boldsymbol{b}},\tilde{\boldsymbol{0}})\in\mathbb{F}_2^{N-(l+M)}$.

\item A polar channel decoder is applied by treating $N-(l+M)$ leaf nodes as frozen and setting them to
$(\tilde{\boldsymbol{b}},\tilde{\boldsymbol{0}})$, while treating the remaining $l+M$ leaf nodes as active. When SCLD is used, the decoder takes $\boldsymbol{\lambda}$ as input and outputs a list of candidate pairs
$(\tilde{\boldsymbol{d}},\tilde{\boldsymbol{p}})\in\mathbb{F}_2^{l+M}$. These candidates are then filtered by the CRC constraint, and only those passing the CRC check are retained. If no candidate passes the CRC check, the hypothesis $l$ is discarded.

\item For each surviving pair $(\tilde{\boldsymbol{d}},\tilde{\boldsymbol{p}})$, a candidate source sequence
$\tilde{\boldsymbol{s}}\in\mathbb{F}_2^K$ is reconstructed from $\tilde{\boldsymbol{d}}$ under the PCL setting $l$.
\end{enumerate}

After all candidates in $\mathcal{L}$ have been tested, the surviving source candidates are compared according to a MAP criterion. For each candidate source sequence
$\tilde{\boldsymbol{s}}$ and its corresponding channel codeword $\tilde{\boldsymbol{x}}$, define the metric
\begin{equation}
R(\tilde{\boldsymbol{s}})
\triangleq
\log P_{\boldsymbol{S}}(\tilde{\boldsymbol{s}})
+
\log W_{\boldsymbol{Y}|\boldsymbol{X}}(\boldsymbol{y}\mid \tilde{\boldsymbol{x}}).
\end{equation}
The final estimate is given by
\begin{equation}
\hat{\boldsymbol{s}}
=
\arg\max_{\tilde{\boldsymbol{s}}} R(\tilde{\boldsymbol{s}}).
\end{equation}
\subsection{Design of the PCL Set}
The PCL set $\mathcal{L}$ plays a fundamental role in the proposed JSCC scheme, as it determines the granularity of rate adaptation and governs the trade-off between source compression efficiency and channel decoding reliability. Consequently, a proper design of $\mathcal{L}$ is crucial to achieving good overall performance.

\subsubsection{Error Analysis and Surrogate Objective}
For a prescribed cardinality $m$, the objective is to select $m$ elements from $\{1,2,\ldots,K\}$ to construct the PCL set $\mathcal{L}=\{l_0,l_1,\ldots,l_{m-1}\}$ with $l_0<l_1<\cdots<l_{m-1}=K$. To guide this design, we first analyze the error performance of the proposed scheme.

Given $\mathcal{L}$, the source space $\mathbb{F}_2^K$ is partitioned into $m$ disjoint classes according to the predicted polar complexity $\overline{L^{\star}}(\boldsymbol{s})$, as shown in~\eqref{SelectedPCL_final}. Specifically, for $i=0,1,\ldots,m-1$, define
\begin{equation}
\mathcal{S}_i 
= 
\left\{ 
\boldsymbol{s} : 
\overline{L^{\star}}(\boldsymbol{s})
\in (l_{i-1}, l_i] 
\right\}
\end{equation}
with $l_{-1}=0$. All sequences in $\mathcal{S}_i$ are assigned the same PCL $l_i$. The decoding error events for $\mathcal{S}_i$ can be decomposed into
\begin{align}
E_i^{(1)} 
&\triangleq 
\left\{ 
\boldsymbol{s} \in \mathcal{S}_i,\ 
\hat{\boldsymbol{s}} \in \mathcal{S}_i,\ 
\hat{\boldsymbol{s}} \neq \boldsymbol{s} 
\right\}, \\
E_i^{(2)} 
&\triangleq 
\left\{ 
\boldsymbol{s} \in \mathcal{S}_i,\ 
\hat{\boldsymbol{s}} \notin \mathcal{S}_i 
\right\}.
\end{align}
Here, $E_i^{(1)}$ corresponds to within-class decoding errors, while $E_i^{(2)}$ represents cross-class errors. Empirical observations show that cross-class errors are significantly less likely than within-class errors. Hence, the FER can be approximated as
\begin{align}
\mathrm{FER}
\approx 
\sum_{i=0}^{m-1} \Pr\{E_i^{(1)}\}.
\end{align}

Let $q(l)$ denote the probability mass function of the predicted compressibility $l$. Then, the probability of $\mathcal{S}_i$ is
\begin{equation}
\Pr\{\mathcal{S}_i\}
=
\sum_{l=l_{i-1}+1}^{l_i} q(l).
\end{equation}
Since all sequences in $\mathcal{S}_i$ use the same PCL $l_i$, they share the same effective channel-code dimension $l_i+M$, where $M$ is the number of CRC bits. Let $\epsilon(l_i+M)$ denote the corresponding decoding error probability. The FER can thus be approximated by
\begin{align}
\mathrm{FER}
\approx
\sum_{i=0}^{m-1} 
\Pr\{\mathcal{S}_i\}\,\epsilon(l_i+M)
=
\sum_{i=0}^{m-1} 
\sum_{l=l_{i-1}+1}^{l_i}
q(l)\,\epsilon(l_i+M).
\label{eq:PCL_design_surrogate}
\end{align}
This expression admits a clear interpretation: all source sequences with predicted compressibility in $(l_{i-1},l_i]$ are grouped together and experience the same channel reliability characterized by $\epsilon(l_i+M)$.

\subsubsection{Optimization Formulation}
Based on the surrogate FER expression in~\eqref{eq:PCL_design_surrogate}, the PCL set design can be formulated as
\begin{align}
\min_{\,l_0 < l_1 < \cdots < l_{m-1} = K}
\quad
\sum_{i=0}^{m-1} 
\sum_{l=l_{i-1}+1}^{l_i}
q(l)\,\epsilon(l_i + M).
\label{eq:PCL_opt_problem}
\end{align}
Here, $q(\cdot)$ can be estimated via Monte Carlo simulation, while $\epsilon(\cdot)$ can be approximated using, for example, density evolution with Gaussian approximation~\cite{Yao2022ABT}\cite{Trifonov2012}.
\subsubsection{Dynamic Programming Approach}
The optimization problem in~\eqref{eq:PCL_opt_problem} can be interpreted as a one-dimensional segmentation problem, where the interval $[1,K]$ is partitioned into $m$ contiguous segments. Each segment $(l_{i-1},l_i]$ incurs a cost equal to the total probability mass within the segment weighted by the decoding error evaluated at its right endpoint.

This structure admits an efficient solution via dynamic programming. Let $F(g,t)$ denote the minimum cost of partitioning the interval $[1,g]$ into $t$ segments. Then, $F(g,t)$ satisfies the recursion
\begin{align}
F(g,t)
=
\min_{g' < g}
\left\{
F(g',t-1)
+
\sum_{l=g'+1}^{g} q(l)\,\epsilon(g+M)
\right\},
\end{align}
with initialization
\begin{align}
F(g,1)=\sum_{l=1}^{g} q(l)\,\epsilon(g+M).
\end{align}
The optimal value is given by $F(K,m)$, and the corresponding partition points $\{l_i\}$ can be recovered via standard backtracking. Therefore, the dynamic programming approach yields the globally optimal PCL set under the surrogate objective in~\eqref{eq:PCL_design_surrogate}, with polynomial computational complexity.

\subsection{Simulation Results}
\subsubsection{Predicted polar complexity}
\begin{figure}
\centering
\includegraphics[width=0.68\linewidth]{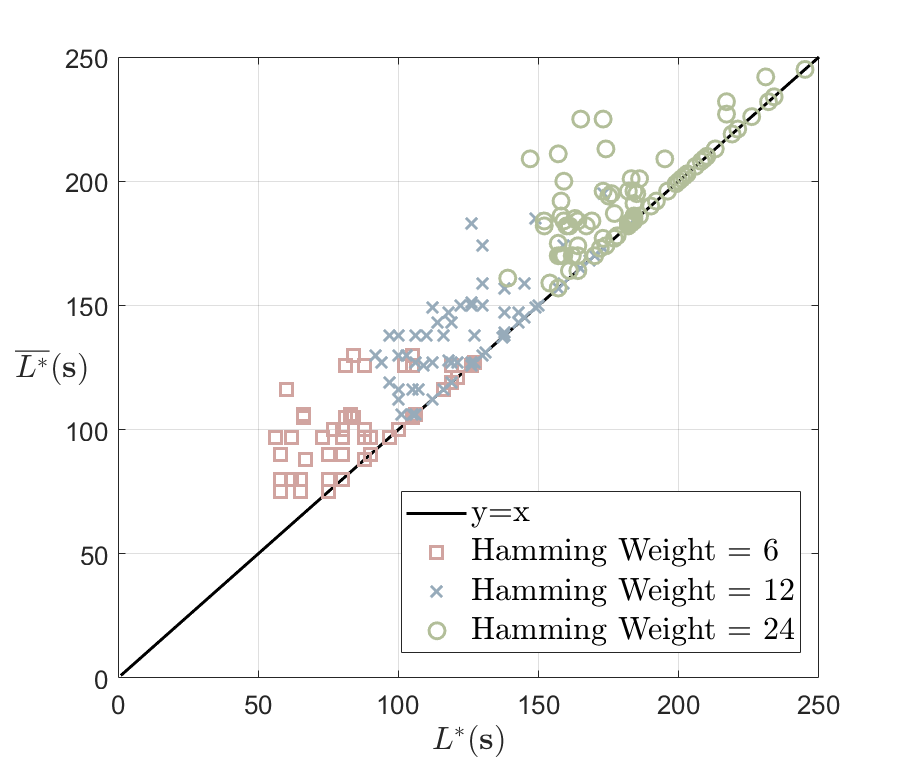}
\caption{Comparison between the predicated polar complexity $\overline{L^*}(\boldsymbol{s})$ and the true polar complexity $L^*(\boldsymbol{s})$ for source length $K=512$ under different Hamming weights.}
\label{PCvsPPC_N512}
\end{figure}
We consider source sequences of length $K=512$ with Hamming weights $6$, $12$, and $24$. For each sequence, both the true polar complexity and its predicted counterpart are evaluated. The results are shown in Fig.~\ref{PCvsPPC_N512}. It is observed that the predicted polar complexity consistently upper-bounds the true polar complexity across all considered Hamming weights, thereby validating~\eqref{eq:Lhat_geq_Lstar}.

Furthermore, the gap between the predicted and true polar complexity remains small, indicating that the proposed predictor is not overly conservative. These results demonstrate that the predictor provides a reliable and efficient estimate of the required PCL. In particular, it ensures sufficient compression for successful decoding while avoiding excessive redundancy, which is crucial for the subsequent JSCC design.

\subsubsection{JSCC Performance}
\begin{figure}[t]
\centering
\includegraphics[width=0.68\linewidth]{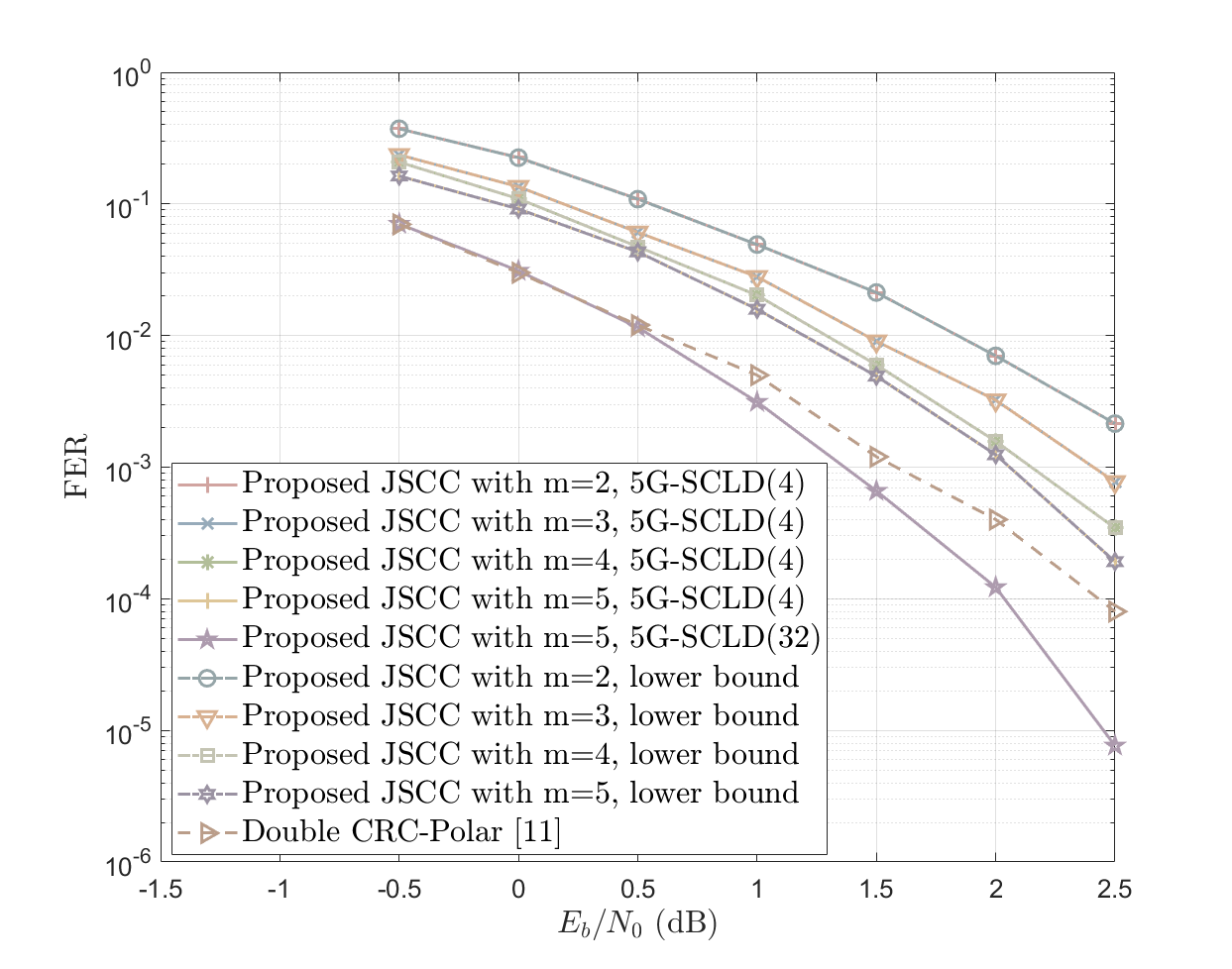}
\caption{FER performance of the proposed JSCC scheme for $(N,K,\theta)=(128,64,0.11)$ under different cardinalities $m$.}
\label{fig:JSCC_small}
\end{figure}

\begin{figure}[t]
\centering
\includegraphics[width=0.68\linewidth]{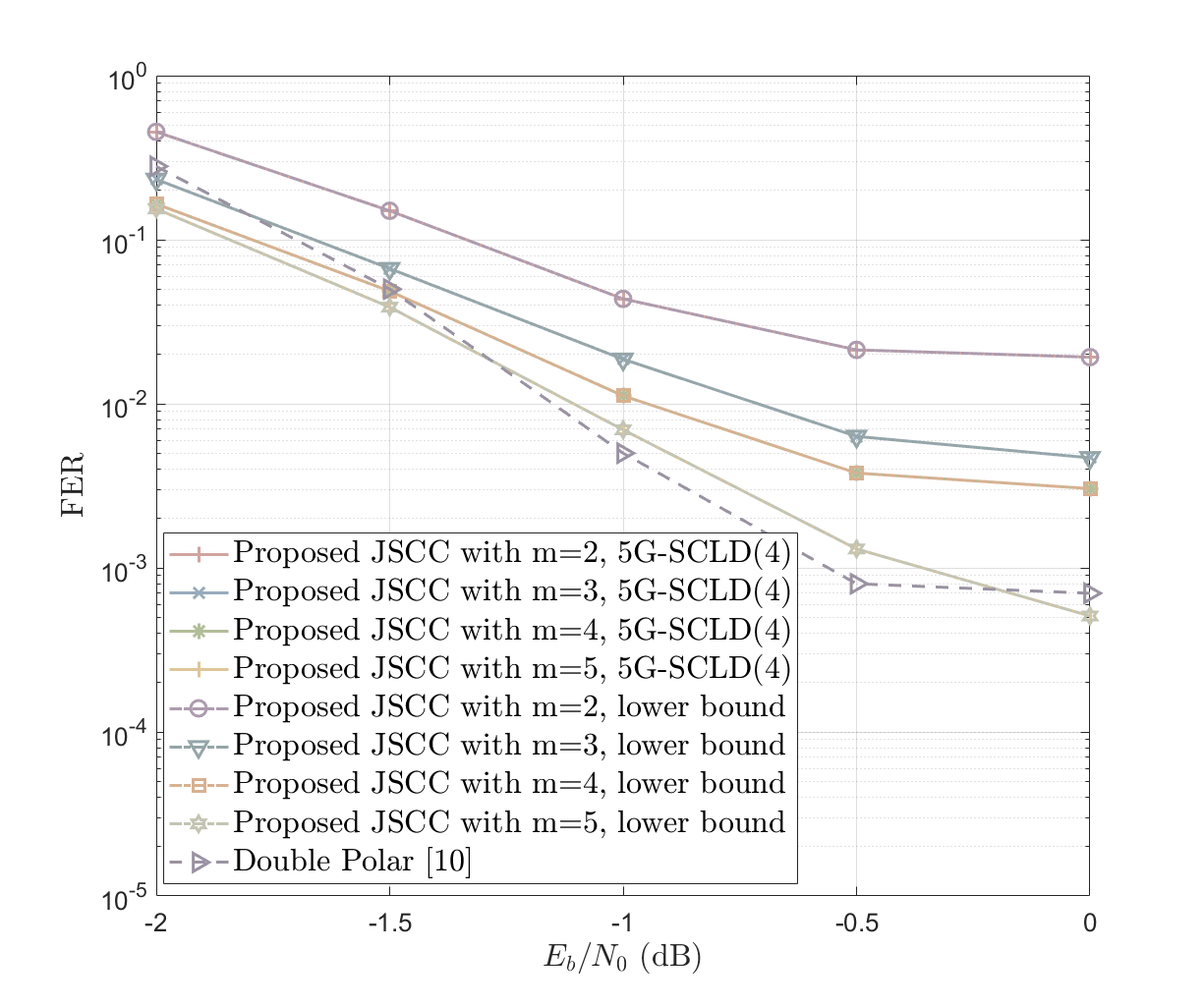}
\caption{FER performance of the proposed JSCC scheme for $(N,K,\theta)=(1024,512,0.07)$ under different cardinalities $m$.}
\label{fig:JSCC_large}
\end{figure}

\begin{table}[t]
\caption{Simulation parameters for evaluating the proposed JSCC scheme}
\label{SimulationParameters}
\centering
\begin{tabular}{@{}ccccc@{}}
\toprule
\begin{tabular}[c]{@{}c@{}}Setting\\ $(N,K,\theta)$\end{tabular} 
& \begin{tabular}[c]{@{}c@{}}$\mathcal{L}$\\ $(m=2)$\end{tabular} 
& \begin{tabular}[c]{@{}c@{}}$\mathcal{L}$\\ $(m=3)$\end{tabular} 
& \begin{tabular}[c]{@{}c@{}}$\mathcal{L}$\\ $(m=4)$\end{tabular} 
& \begin{tabular}[c]{@{}c@{}}$\mathcal{L}$\\ $(m=5)$\end{tabular} \\ 
\midrule
$(128,64,0.11)$   
& $\{47,64\}$ 
& $\{42,51,64\}$ 
& $\{39,47,54,64\}$ 
& $\{38,44,49,55,64\}$ \\
$(1024,512,0.07)$ 
& $\{310,512\}$ 
& $\{282,334,512\}$ 
& $\{270,307,344,512\}$ 
& $\{270,301,326,358,512\}$ \\ 
\bottomrule
\end{tabular}
\end{table}

We evaluate the proposed JSCC scheme under two settings, with the corresponding optimized PCL sets for different cardinalities summarized in Table~\ref{SimulationParameters}. In both cases, we set the number of CRC bits $M=8$ and adopt the CRC generator polynomial $x^8 + x^7 + x^6 + x^4 + x^2 + 1$. 

The FER performance is shown in Fig.~\ref{fig:JSCC_small} and Fig.~\ref{fig:JSCC_large}, where ``5G-SCLD$(x)$'' denotes 5G-sequence construction with SCLD of list size $x$ for channel coding. A lower bound is also included, corresponding to the case where the true PCL selected at the encoder is revealed to the decoder. 

As observed, the FER decreases monotonically with the PCL-set cardinality $m$ in both scenarios. This behavior is attributed to the improved adaptation to sequence-dependent polar complexity enabled by a finer PCL discretization. Moreover, the performance curves closely approach the lower bound, indicating near-optimal performance.

We further compare the proposed scheme with existing polar-based JSCC schemes, including the double-CRC-polar scheme in~\cite{Zheng2021DoubleCRCPolar} for the short-blocklength regime and the double-polar scheme with joint SCLD in~\cite{Dong2022DoublePolarJSCL} for the long-blocklength regime. The proposed scheme achieves competitive performance across both regimes.

Overall, the proposed scheme offers a flexible performance--complexity trade-off governed by the PCL-set cardinality $m$, and demonstrates consistent effectiveness over both short and long blocklengths.

\section{Conclusion}\label{sec6}
In this paper, we have proposed the concept of polar complexity, which can be efficiently computed via a bisection search method. To facilitate the rapid estimation of polar complexity, we have further defined a critical set, whose cardinality serves as the upper bound on the polar complexity.

Based on polar complexity and the corresponding polar-induced description, we have presented a polar-based two-stage source coding scheme for BMSs. Each source sequence is represented by a fixed-length header specifying its polar complexity, together with a variable-length polar-induced description. This construction is strictly lossless and prefix-free, and is further shown to be entropy-achieving when the source length is a power of two. Simulation results demonstrate its competitive finite-length compression performance and robustness with respect to the source distribution.

We have further developed an adaptive double-polar JSCC scheme that unifies the proposed polar-based two-stage source coding and polar channel coding within a single framework. The introduction of a finite PCL set enables a controllable tradeoff between adaptability and complexity, while allowing the encoder to match the predicted compressibility of each source realization with the protection capability of the channel code. The PCL set design has also been formulated as an optimization problem and solved efficiently via dynamic programming. Simulation results demonstrate that the proposed adaptive double-polar JSCC scheme achieves superior error-rate performance compared to existing polar-based JSCC baselines.

\ifCLASSOPTIONcaptionsoff
\newpage
\fi
\bibliographystyle{IEEEtran}
\bibliography{IEEEabrv,RefYaoxym}
\end{document}